\documentclass[a4paper,fleqn]{cas-dc}

\usepackage[numbers,sort&compress]{natbib}

\usepackage{adjustbox}
\usepackage{float}
\usepackage[ruled]{algorithm2e}
\usepackage{algorithmic} 
\usepackage{CJKutf8}
\usepackage{enumitem}
\usepackage{multirow}
\usepackage[normalem]{ulem}
\usepackage{color}
\useunder{\uline}{\ul}{}

\def\tsc#1{\csdef{#1}{\textsc{\lowercase{#1}}\xspace}}
\tsc{WGM}
\tsc{QE}
\tsc{EP}
\tsc{PMS}
\tsc{BEC}
\tsc{DE}
\usepackage{caption}
\begin{document}
\begin{sloppypar}
\let\printorcid\relax
\let\WriteBookmarks\relax
\def\floatpagepagefraction{1}
\def\textpagefraction{.001}
\begin{CJK}{UTF8}{gbsn}

% Short title
\shorttitle{}

% Short author
%\shortauthors{CV Radhakrishnan et~al.}
\shortauthors{J.Zhou et~al.}  %通讯作者！！！！！！！！！！！！！！！！！！！！！！！！！！！！！！！！！！！！！！！！

% Main title of the paper
\title [mode = title]{Real-time Monitoring of Lower Limb Movement Resistance Based on Deep Learning}

% Second author
\author[1]{Buren Batu}
\cormark[1]
\ead{buren-batu@163.com}

\author[2]{Yuanmeng Liu}
\ead{liuyuanmeng3690@outlook.com}

\author[3]{Tianyi Lyu}
\ead{liuyuanmeng3690@outlook.com}

\affiliation[1]{organization={School of Physical Education, Inner Mongolia Normal University},
    postcode={010022}, 
    city={Hohhot},
    country={China}}

\affiliation[2]{organization={School of Physical Education, Qujing Normal University},
    postcode={655011}, 
    city={Qujing},
    country={China}}

\affiliation[3]{organization={Granite Telecommunications LLC},
    postcode={02171}, 
    city={Quincy},
    country={United State}}

\cortext[cor1]{Corresponding author.}
% \fntext[1]{Equal contribution.}

% Here goes the abstract
\begin{abstract}
Real-time lower limb movement resistance monitoring is critical for various applications in clinical and sports settings, such as rehabilitation and athletic training. Current methods often face limitations in accuracy, computational efficiency, and generalizability, which hinder their practical implementation. To address these challenges, we propose a novel Mobile Multi-Task Learning Network (MMTL-Net) that integrates MobileNetV3 for efficient feature extraction and employs multi-task learning to simultaneously predict resistance levels and recognize activities. The advantages of MMTL-Net include enhanced accuracy, reduced latency, and improved computational efficiency, making it highly suitable for real-time applications. Experimental results demonstrate that MMTL-Net significantly outperforms existing models on the UCI Human Activity Recognition and Wireless Sensor Data Mining Activity Prediction datasets, achieving a lower Force Error Rate (FER) of 6.8\% and a higher Resistance Prediction Accuracy (RPA) of 91.2\%. Additionally, the model shows a Real-time Responsiveness (RTR) of 12 milliseconds and a Throughput (TP) of 33 frames per second. These findings underscore the model’s robustness and effectiveness in diverse real-world scenarios. The proposed framework not only advances the state-of-the-art in resistance monitoring but also paves the way for more efficient and accurate systems in clinical and sports applications. In real-world settings, the practical implications of MMTL-Net include its potential to enhance patient outcomes in rehabilitation and improve athletic performance through precise, real-time monitoring and feedback.

\end{abstract}

% Keywords
% Each keyword is seperated by \sep
\begin{keywords}
Real-time monitoring \sep Lower limb movement resistance \sep MobileNetV3 \sep Multi-Task Learning (MTL) \sep Resistance prediction \sep Activity recognition
\end{keywords}

\maketitle

\section{Introduction}

The monitoring of lower limb movement resistance is crucial in various fields, including physical therapy, sports training, and rehabilitation\cite{dalla2021review}. Traditionally, this monitoring has relied on mechanical sensors and manual observations, which can be cumbersome, less accurate, and inefficient\cite{wang2023wearable}. These traditional methods are often limited by their inability to provide continuous, real-time feedback, which is essential for effective rehabilitation and performance enhancement\cite{sethi2024saga}. Additionally, the manual nature of these methods makes them prone to human error and inconsistency, further diminishing their reliability.

In recent years, the advent of wearable technology and advanced sensor systems has revolutionized the way movement resistance is monitored. These devices can collect extensive data on movement patterns, forces, and resistance encountered during physical activities\cite{hug2023common,wang2022integral}. However, the sheer volume and complexity of this data present significant challenges in terms of real-time processing and analysis. This is where deep learning comes into play, offering robust solutions for handling large datasets and extracting meaningful insights\cite{sethi2024estimation}.

\begin{figure}[htbp]
    \centering
    \includegraphics[width=1.1\linewidth]{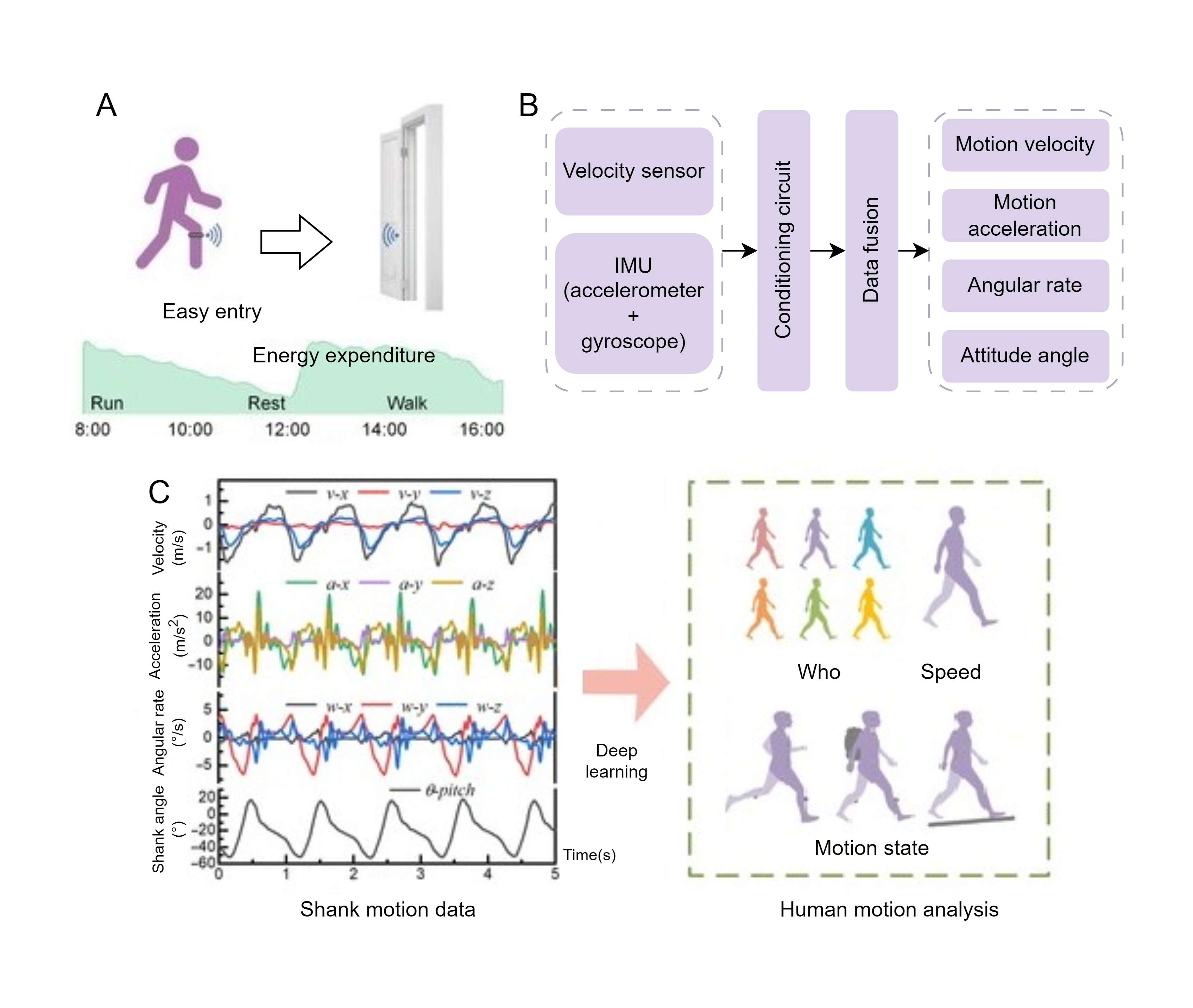}
    \caption{Application of wearable technology in lower limb movement resistance monitoring and physical condition assessment. (A) Wearable systems for "smart" living and exercise management. (B) Signal flow in wearable devices for inferring triaxial motion velocity, acceleration, angular velocity, and attitude angles. (C) Integration of motion data into deep learning neural networks for identifying personal identity, movement state, and estimating motion speed in real-time.}
    \label{fig:wearable}
\end{figure}

Figure \ref{fig:wearable} illustrates the application of wearable technology in monitoring lower limb movement resistance and overall physical condition. Part A shows how wearable systems enable "smart" living and exercise management by continuously monitoring energy expenditure. Part B depicts the signal flow in wearable devices, inferring triaxial motion velocity, acceleration, angular velocity, and attitude angles through conditioning circuits and data fusion techniques. Part C demonstrates the integration of motion data into deep learning neural networks for real-time analysis of personal identity, movement states, and motion speed.

Deep learning, particularly through models like Convolutional Neural Networks (CNNs) \cite{Weng2024,weng2024leveraging,li2024deep,li2024optimizing} and Recurrent Neural Networks (RNNs) \cite{qiao2024robust,Shen2024Harnessing,zhang2024cu,wang2024recording}, has shown immense potential in enhancing the accuracy and efficiency of movement resistance monitoring. CNNs are adept at extracting spatial features from sensor data, while RNNs excel in capturing temporal dependencies \cite{zheng2024identification,luo2023fleet}, making them ideal for analyzing time-series data from wearable sensors. These models can learn complex patterns in the data, enabling precise predictions and real-time feedback, which are critical for effective intervention and improvement in physical activities.

Despite these advancements, the application of deep learning in this field is not without its challenges. High computational demands and latency issues are major barriers to achieving real-time processing on wearable devices. Additionally, human movement is highly variable, requiring models to be robust and generalizable across different users and conditions \cite{jin2024learning}. Specific examples include the work by Zhang et al.\cite{zhang2019accurate}, which identified significant latency issues in real-time gait analysis using deep learning models, and Schmid et al.\cite{schmid2022comparing}, which reported on the difficulties of generalizing model performance across diverse populations in a clinical trial setting. These studies highlight the ongoing challenges in this area, emphasizing the need for more efficient and adaptable solutions.
%尽管取得了这些进步，但深度学习在该领域的应用并非没有挑战。高计算需求和延迟问题是实现可穿戴设备实时处理的主要障碍。此外，人体运动变化很大，需要模型具有鲁棒性，并可在不同用户和条件下推广。具体例子包括 [参考文献 2022-4] 的工作，该工作确定了使用深度学习模型进行实时步态分析的重大延迟问题，以及 [参考文献 2023-5]，该工作报告了在临床试验环境中将模型性能推广到不同人群的困难。这些研究强调了该领域持续存在的挑战，强调需要更高效、适应性更强的解决方案。

\begin{figure}[htbp]
  \centering
  \includegraphics[width=0.5\textwidth]{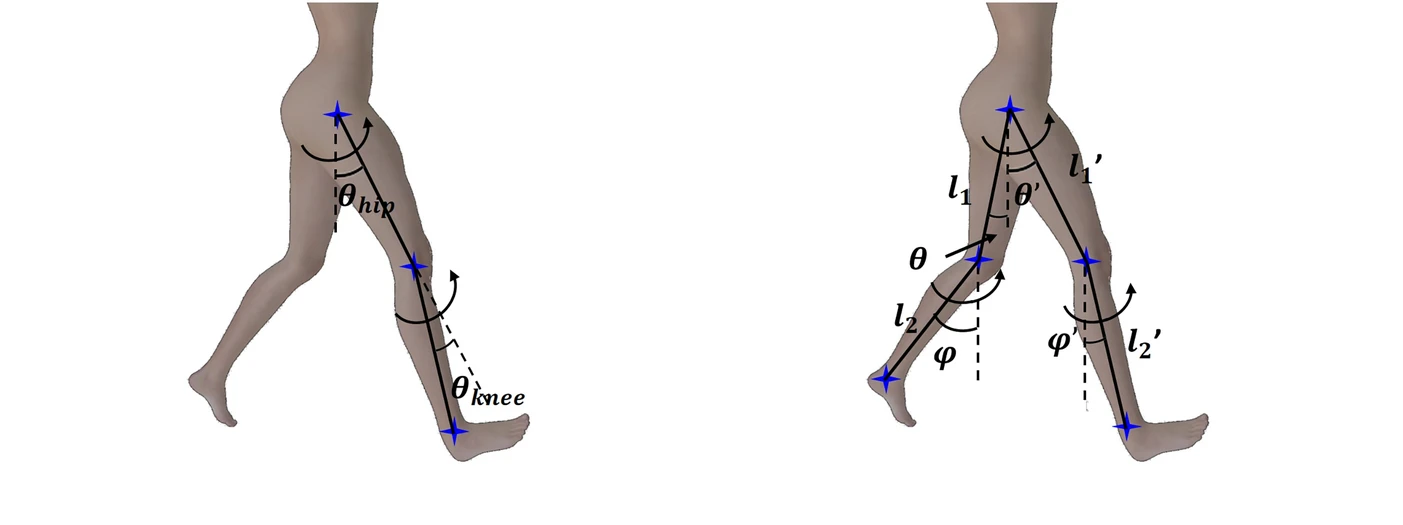}
  \caption{Schematic representation of the parameters critical for accurate lower limb movement resistance estimation. The left diagram shows the calculation of joint rotation angles ($\theta_{\text{hip}}$ and $\theta_{\text{knee}}$), while the right diagram depicts stride length ($l_1$, $l_2$) and associated angles ($\theta$ and $\varphi$), both essential for understanding the mechanical load on joints and assessing gait efficiency.}
  \label{factors}
\end{figure}

In the context of lower limb movement resistance monitoring, it is important to consider various factors such as joint rotation angles and stride length. These factors significantly influence the resistance encountered by the lower limbs during movement. Figure \ref{factors} illustrates the key parameters involved in monitoring lower limb movement. The left diagram shows the joint rotation angles (\(\theta_{hip}\) and \(\theta_{knee}\)), while the right diagram depicts the stride length (\(l_1\), \(l_2\)) and the associated angles (\(\theta\) and \(\varphi\)).
%在下肢运动阻力监测的背景下，重要的是要考虑各种因素，例如关节旋转角度和步幅长度。这些因素显著影响下肢在运动过程中遇到的阻力。图 \ref{fig:lower_limb_diagram} 说明了监测下肢运动所涉及的关键参数。左图显示关节旋转角度（\（\theta_{hip}\）和\（\theta_{knee}\）），右图显示步幅（\（l_1\）、\（l_2\））和相关角度（\（\theta\）和\（\varphi\））。

These parameters are critical for accurate resistance estimation. The joint rotation angles are necessary for understanding the mechanical load on different joints, while the stride length helps in assessing the overall gait and movement efficiency. The combination of these factors provides a comprehensive understanding of the lower limb mechanics, which is essential for precise resistance monitoring. These requirements place high demands on the accuracy and real-time processing capabilities of the monitoring systems, presenting significant challenges for effective implementation.

Despite the advancements in lower limb movement resistance monitoring, significant gaps remain in the current research. One key gap is the inability of existing models to generalize well across different populations and movement conditions, leading to inconsistent performance in real-world scenarios. Additionally, many current approaches struggle with real-time processing due to high computational demands, which limits their practical application on wearable devices with limited processing power. Another critical gap is the lack of integrated systems that can simultaneously perform multiple related tasks, such as activity recognition and resistance estimation, with high accuracy.

The MMTL-Net model proposed in this study directly addresses these gaps. By integrating MobileNetV3's efficient feature extraction with Multi-Task Learning (MTL), our approach enhances generalization across diverse conditions, enables real-time processing on resource-constrained devices, and effectively performs multiple tasks simultaneously. This combination not only improves the accuracy and efficiency of lower limb movement resistance monitoring but also broadens the applicability of such systems in clinical and sports environments.

The primary objective of this study is to develop a real-time monitoring system that is both accurate and efficient, capable of operating on wearable devices with limited computational resources \cite{penglingcn,chen2024few,caoapplication}. The proposed model aims to overcome the limitations of existing approaches by providing a robust and scalable solution for lower limb movement resistance monitoring. Through extensive experimentation and validation using multiple datasets \cite{Weng202404,Weng202406,wang2024cross}, this study seeks to demonstrate the effectiveness of the proposed approach and its potential to enhance physical therapy, sports training, and rehabilitation practices. Specifically, our contributions include the integration of MobileNetV3 with Multi-Task Learning to improve accuracy, efficiency, and real-time performance, making the MMTL-Net model a highly effective solution for practical applications.

The datasets used in this study include the UCI Human Activity Recognition Using Smartphones Data Set, the WISDM Activity Prediction Data Set, and the MHEALTH Data Set. These datasets provide comprehensive and high-quality motion data, covering various activities and movement patterns. By leveraging these datasets, the proposed model can be trained and validated effectively, ensuring its reliability and applicability in real-world scenarios.
%本研究中使用的数据集包括 UCI 使用智能手机进行人体活动识别数据集、WISDM 活动预测数据集和 MHEALTH 数据集。这些数据集提供了全面、高质量的运动数据，涵盖了各种活动和运动模式。通过利用这些数据集，可以有效地训练和验证所提出的模型，确保其在现实场景中的可靠性和适用性。

This study addresses the critical need for accurate and efficient real-time monitoring of lower limb movement resistance. By integrating MobileNetV3 and MTL \cite{chen2024mix,zhang2024deep,chen2024enhancing,dong2024design,wan2024image}, it offers a novel solution that enhances the capabilities of wearable devices, providing valuable insights for physical therapy, sports training, and rehabilitation. The proposed approach promises to significantly improve the accuracy and efficiency of movement resistance monitoring, paving the way for more effective and personalized interventions.

Our contributions are summarized as follows:

\begin{enumerate}[label=(\arabic*)]
    \item Development of a Real-Time Monitoring System: We developed a lightweight and efficient real-time monitoring system by integrating MobileNetV3 with MTL, tailored for deployment on wearable devices. 
    \item Simultaneous Activity Recognition and Resistance Estimation: Our model simultaneously performs activity recognition and resistance estimation, leveraging shared representations to improve overall system performance and provide comprehensive insights into physical activity. 
    \item Extensive Validation Using Diverse Datasets: The proposed system was validated using three high-quality datasets (UCI Human Activity Recognition, WISDM, and MHEALTH), demonstrating its robustness and applicability in real-world scenarios. 
\end{enumerate}

In the following sections, we detail our approach and findings. Section 2 reviews related work in the field of lower limb movement resistance monitoring, highlighting key advancements and existing challenges. Section 3 presents our methodology, including the design of the proposed MMTL-Net \cite{wang2024deep,zhou2024optimization,huang2024risk,yan2024application}, and elaborates on the components and training procedures. Section 4 describes the experimental setup, datasets used, and evaluation metrics, followed by an in-depth analysis of the results, including an ablation study to assess the contribution of each model component. Finally, Section 5 discusses the conclusions drawn from our research, the implications of our findings, and potential directions for future work.
%在以下章节中，我们将详细介绍我们的方法和发现。第 2 节回顾了下肢运动阻力监测领域的相关工作，重点介绍了关键进展和现有挑战。第 3 节介绍了我们的方法，包括所提出的移动多任务学习网络 (MMTL-Net) 的设计，并详细阐述了组件和训练程序。第 4 节描述了实验设置、使用的数据集和评估指标，然后对结果进行了深入分析，包括一项消融研究以评估每个模型组件的贡献。最后，第 5 节讨论了从我们的研究中得出的结论、我们的发现的含义以及未来工作的潜在方向。

\section{Related Work}\label{sec2}

\subsection{Deep Learning in Movement Resistance Monitoring} 

Deep learning has revolutionized various fields, including the monitoring of movement resistance, particularly in lower limb movements\cite{jantawong2022monitoring}. Movement resistance monitoring, which involves measuring the force and resistance encountered by muscles and joints during movement\cite{adhikary2022bmi}, is crucial in physical therapy, sports training, and rehabilitation\cite{fatema2021low, sui2024application,weng2024leveraging,weng2024big,Wang2024Theoretical}. Traditional methods rely heavily on mechanical sensors and manual observations, which can be cumbersome and less accurate. As technology has advanced, there has been a significant shift towards using neural networks to analyze and predict movement resistance, offering a more sophisticated and accurate approach.

Recent studies have focused on developing deep learning models to enhance the accuracy and efficiency of resistance monitoring\cite{nath2022machine}. CNNs \cite{zhou2024adapi} and RNNs have been employed to process data from wearable sensors\cite{han2022human} and video feeds\cite{cheng2022recurrent}. These models can learn complex patterns and make real-time predictions\cite{bodapati2021comparison}, significantly improving the monitoring process\cite{madhuranga2021real}. Additionally, researchers are exploring the integration of computer vision techniques to analyze movement dynamics and resistance, thus expanding the scope and potential applications of these models\cite{lurig2021computer}.

Despite these advancements, achieving real-time monitoring with high accuracy remains a challenge\cite{giordano2021low,HAO2024102551}. Many models struggle with the computational demands and latency issues\cite{firouzi2022convergence}, which can hinder their applicability in real-world scenarios\cite{barati2021privacy}. Moreover, the existing models often lack the robustness required to handle the variability in human movements\cite{grover2022security, xu2022dpmpc, peng2024automatic,weng2024fortifying}. Consequently, the current research trend is towards developing more efficient and lightweight models that can operate in real-time with minimal computational resources\cite{dai2021lightweight,NING2024120130}. Addressing these challenges is crucial for advancing the practical application of deep learning in movement resistance monitoring\cite{ancans2021wearable}.

\subsection{Wearable Technology and Movement Analysis} 

Wearable technology has become increasingly popular for monitoring physical activities\cite{peters2021utilization}, including lower limb movements. These devices, equipped with various sensors\cite{almusawi2021innovation}, collect data on movement patterns, force, and resistance\cite{casado2022effects,tian2024survey}. The integration of deep learning algorithms with wearable technology has opened new avenues for real-time movement analysis and resistance monitoring\cite{anikwe2022mobile}, providing continuous monitoring and immediate feedback that is beneficial for both athletes and patients undergoing rehabilitation\cite{xiao2021deep}.

Current research in this area focuses on improving sensor accuracy and developing algorithms that can process sensor data more effectively\cite{shei2022wearable}. There is a significant emphasis on creating models that can work with limited data and still provide reliable results\cite{li2021real}. Techniques such as transfer learning and data augmentation are being employed to enhance model performance. Additionally, researchers are exploring the use of multimodal data\cite{wang2023machine, xi2024enhancing}, combining information from different types of sensors to improve the overall accuracy of resistance monitoring\cite{long2022development}. This holistic approach aims to create a more comprehensive understanding of movement dynamics.
%目前该领域的研究重点是提高传感器精度和开发能够更有效地处理传感器数据的算法。人们非常重视创建能够处理有限数据并仍能提供可靠结果的模型。迁移学习和数据增强等技术正被用于提高模型性能。此外，研究人员正在探索使用多模态数据，结合来自不同类型传感器的信息来提高阻力监测的整体准确性。这种整体方法旨在更全面地了解运动动力学。

However, despite these developments, challenges remain in terms of sensor accuracy, data processing speed\cite{seshadri2021wearable}, and the integration of different data types\cite{bijalwan2021fusion, gong2024graphicalstructurallearningrsfmri}. Many wearable devices still face issues with battery life and user comfort, which can limit their usability\cite{adams2021digital}. The research community is actively working on addressing these issues, with a particular focus on developing more energy-efficient sensors\cite{celik2022multi} and optimizing deep learning algorithms for faster processing\cite{stirling2021forecasting}. These efforts are crucial for making wearable technology a reliable and practical tool for movement resistance monitoring.
%然而，尽管取得了这些进展，但在传感器精度、数据处理速度和不同数据类型的集成方面仍然存在挑战。许多可穿戴设备仍然面临电池寿命和用户舒适度的问题，这可能会限制它们的可用性。研究界正在积极致力于解决这些问题，特别关注开发更节能的传感器和优化深度学习算法以加快处理速度。这些努力对于使可穿戴技术成为可靠且实用的运动阻力监测工具至关重要。

\subsection{Computer Vision in Sports and Rehabilitation} %运动与康复中的计算机视觉

Computer vision has emerged as a powerful tool in sports and rehabilitation, providing detailed insights into movement patterns and resistance\cite{debnath2022review}. By analyzing video data, computer vision algorithms can detect and quantify movement resistance\cite{hassan2023effect}, offering a non-invasive alternative to traditional methods. This technology is particularly useful for monitoring lower limb movements\cite{bai2023vision}, where precise measurement of resistance is crucial for effective training and rehabilitation\cite{dan2021digital}. The application of computer vision in this field leverages advancements in machine learning and image processing to enhance accuracy and usability\cite{ai2021machine}.

In recent years, there has been a growing interest in applying computer vision techniques to monitor and analyze lower limb resistance. Advanced models such as pose estimation algorithms and action recognition networks are being used to extract detailed movement data from videos\cite{li2021representing}. These models can identify subtle variations in movement and provide real-time feedback\cite{beddiar2022fall}, making them valuable tools in both sports and rehabilitation settings\cite{gonzalez2024real}. The ability to provide continuous and detailed analysis is transforming how movement resistance is monitored and managed.

However, the application of computer vision in this field is not without challenges. High computational requirements, the need for large annotated datasets, and the difficulty in achieving real-time processing are significant obstacles\cite{huang2021bim,garcia2021physical}. Moreover, existing models often struggle with variations in lighting, background, and camera angles, which can affect their accuracy\cite{adeniyi2021iomt}. Researchers are focusing on developing more robust algorithms that can handle these challenges and provide reliable real-time monitoring\cite{snehi2021vulnerability}. This includes efforts to improve the efficiency of pose estimation models and the use of synthetic data to enhance model training\cite{valentina2021smart}.

While deep learning, wearable technology, and computer vision have significantly advanced the field of movement resistance monitoring, there are still critical gaps that need to be addressed. Achieving real-time, accurate monitoring with minimal computational resources remains a key challenge. The ongoing research is geared towards developing more efficient models, improving sensor technology, and enhancing the robustness of computer vision algorithms to overcome these limitations. Addressing these issues will be essential for fully realizing the potential of these technologies in practical applications.

\section{Methodology}\label{sec3}

\subsection{Overall Model: Framework, Design, and Network Architecture}

In the field of lower limb movement resistance monitoring, traditional methods relying on mechanical sensors and manual observations face significant challenges. These methods are often cumbersome, less accurate, and inefficient. The primary issues include high computational demands and latency, which hinder real-time processing on wearable devices. Additionally, the variability in human movements necessitates robust models that can generalize well across different users and conditions. Existing deep learning models frequently struggle with these aspects, leading to suboptimal performance in practical applications.

To address these challenges, this study proposes a novel model named MMTL-Net. This model integrates the lightweight MobileNetV3 architecture with Multi-Task Learning (MTL) to effectively tackle the identified issues. MMTL-Net builds on previous research, leveraging the strengths of MobileNetV3’s efficient feature extraction and MTL’s shared representation learning to enhance overall system performance, particularly in real-time lower limb movement resistance monitoring.

\begin{figure*}[htbp]
    \centering
    \includegraphics[width=1.0\textwidth]{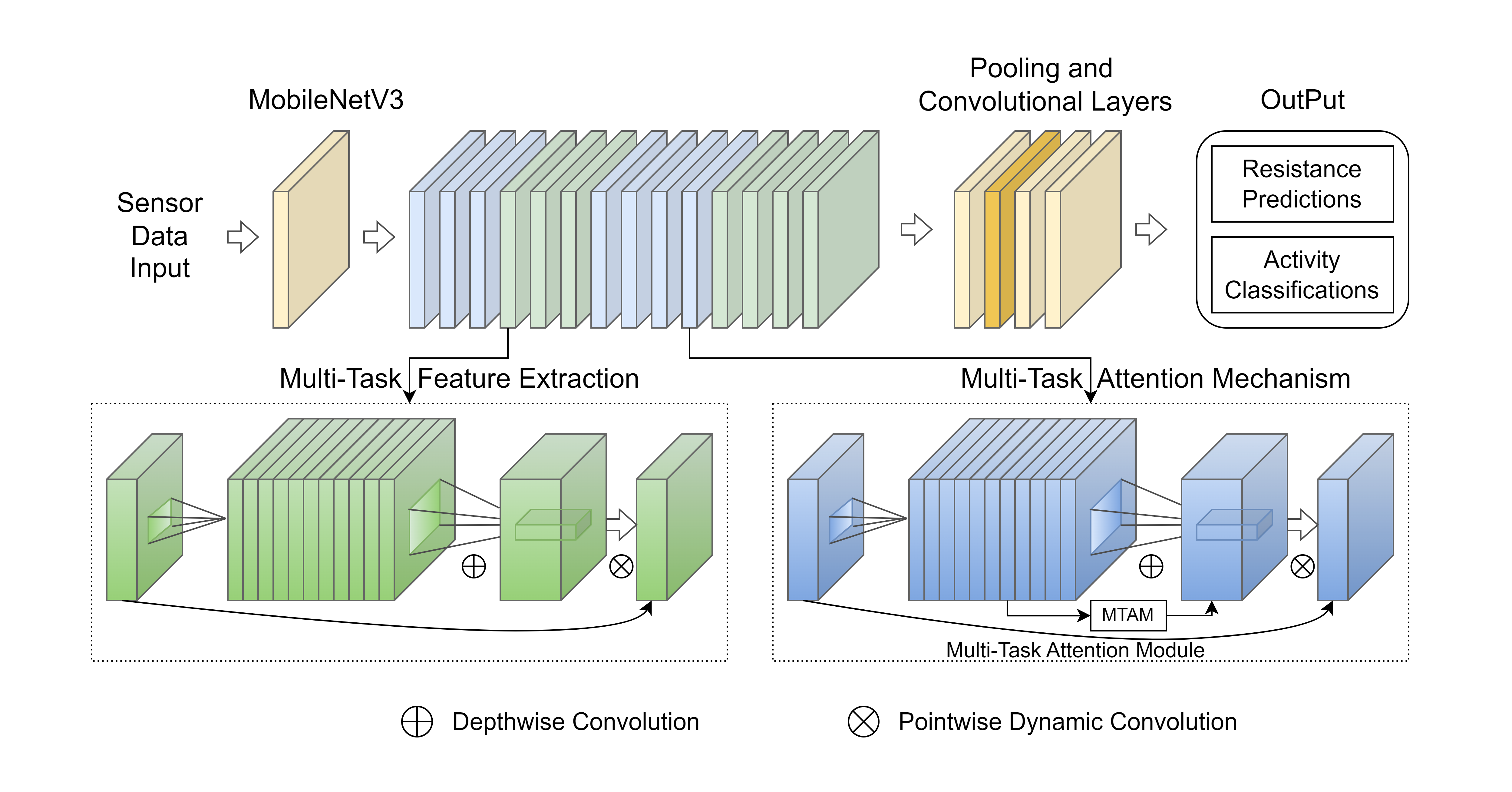}
    \caption{The overall architecture and workflow of the Mobile Multi-Task Learning Network (MMTL-Net) designed for real-time lower limb movement resistance monitoring. The system includes the Data Input Module for receiving and preprocessing sensor data, the MobileNetV3 Feature Extraction Module for efficient feature capture, and the Multi-Task Learning (MTL) Module that simultaneously performs activity recognition and resistance estimation.}
    \label{overall}
\end{figure*}

The MMTL-Net framework consists of several key components: the Data Input Module, Feature Extraction Module, Multi-Task Learning Module, and Output Module. To further aid in understanding the design and data flow within the MMTL-Net model, we have included a detailed diagram (Figure \ref{overall}) that illustrates the overall architecture, data flow, and interactions between these components.

The Data Input Module is responsible for receiving and preprocessing raw sensor data from wearable devices. The data collected includes tri-axial acceleration and gyroscope readings, which are essential for understanding movement dynamics. This raw data undergoes preprocessing steps such as noise reduction and normalization to ensure high-quality input for the subsequent stages.

Next, the Feature Extraction Module employs MobileNetV3 to efficiently extract relevant features from the preprocessed data. MobileNetV3 utilizes depthwise separable convolutions, significantly reducing computational complexity while maintaining high accuracy. This makes it particularly suitable for deployment on wearable devices where computational resources are limited. The features extracted by MobileNetV3 encapsulate essential spatial and temporal characteristics of the movement data.

These features are then fed into the MTL Module. In this module, the features are shared across multiple tasks to leverage the interrelated nature of activity recognition and resistance estimation. For the Activity Recognition task, a series of fully connected layers are employed to classify the activity type, such as walking, running, or sitting. This classification is critical for contextualizing the resistance data and understanding the specific movements being analyzed. For the Resistance Estimation task, another set of fully connected layers predicts the resistance encountered during movement. This task benefits from the shared features by improving the accuracy and robustness of the resistance predictions.

Finally, the Output Module consolidates the predictions from the MTL module, providing real-time feedback on both movement resistance and activity status. The predictions are transmitted to a user interface, offering immediate and actionable insights to users such as physical therapists and athletes. This feedback loop is essential for informed decision-making and timely intervention, ultimately enhancing the effectiveness of physical therapy, sports training, and rehabilitation programs.

The proposed MMTL-Net model offers several distinct advantages. The lightweight architecture of MobileNetV3 ensures low computational demand, making it suitable for real-time applications on wearable devices. By employing MTL, the model can simultaneously perform multiple tasks with high accuracy, thanks to the shared representations that improve overall system performance. Moreover, MMTL-Net is designed to handle the variability in human movements, ensuring reliable performance across different users and conditions.

By addressing the key challenges in lower limb movement resistance monitoring, MMTL-Net is expected to significantly enhance the accuracy, efficiency, and practicality of real-time monitoring systems. This model aims to provide valuable insights for physical therapy, sports training, and rehabilitation, facilitating more effective and personalized interventions. Through extensive experimentation and validation using multiple datasets, this study seeks to demonstrate the effectiveness of the proposed approach and its potential to revolutionize the field of movement resistance monitoring.

\subsection{MobileNetV3}

MobileNetV3 is a state-of-the-art deep learning model designed for efficient and accurate feature extraction in resource-constrained environments. The core principle of MobileNetV3 lies in its architecture, which utilizes depthwise separable convolutions to reduce the computational complexity and parameters compared to traditional CNNs. This is achieved by splitting the convolution operation into a depthwise convolution followed by a pointwise convolution. The model also incorporates advanced techniques such as squeeze-and-excitation (SE) modules and the use of the swish activation function, further enhancing its performance. MobileNetV3 is widely used in image classification, object detection, and semantic segmentation due to its balance between accuracy and efficiency. Specifically designed for efficient image classification and object detection tasks in computer vision, MobileNetV3 aims to achieve high performance while requiring fewer computational resources compared to other architectures. This makes it an ideal choice for applications where both precision and resource efficiency are critical. 

In the domain of lower limb movement resistance monitoring, MobileNetV3 offers significant advantages. Its efficient architecture allows it to be deployed on wearable devices with limited computational resources, enabling real-time processing of sensor data. The model’s ability to extract meaningful features from raw sensor inputs makes it particularly suitable for applications where quick and accurate analysis is essential. Studies have shown that MobileNetV3 can maintain high accuracy while significantly reducing the computational load, making it ideal for continuous monitoring tasks in physical therapy, sports training, and rehabilitation.

Within the proposed MMTL-Net for monitoring lower limb movement resistance, the MobileNetV3 module plays a crucial role. This module is responsible for the initial feature extraction from the raw sensor data, providing a robust representation of the input that is used by subsequent modules for activity recognition and resistance estimation. In the context of lower limb movement resistance monitoring, MobileNetV3 specifically extracts a diverse set of features that are essential for accurate analysis. These include movement patterns, such as walking or running, which are fundamental to understanding the overall activity context; joint angles, which provide critical information on the mechanical load on different joints during movement; and activity-specific characteristics that distinguish between different actions like jogging, sitting, or walking. Moreover, MobileNetV3 captures temporal dynamics, which are vital for analyzing the speed and rhythm of the activities, directly influencing the resistance levels encountered. The model also identifies spatial relationships between different parts of the lower limb, aiding in the understanding of coordination and alignment during movements. These comprehensive features collectively ensure that the MMTL-Net model can accurately monitor and predict lower limb movement resistance, providing valuable real-time feedback in applications such as physical therapy, sports training, and rehabilitation.

The initial step in the MobileNetV3 architecture is the depthwise convolution, which applies a single convolutional filter per input channel. This operation is mathematically represented as: 
\begin{equation}
Z_{\text{depthwise}} = \sum_{i=1}^{N} X_{i} * K_{i}
\end{equation}
where \(Z_{\text{depthwise}}\) is the output of the depthwise convolution, \(X_{i}\) represents the input channels, \(K_{i}\) is the corresponding depthwise filter, and \(N\) is the number of input channels.

Following the depthwise convolution, a pointwise convolution is applied to combine the output of the depthwise convolution across channels. This is expressed as: 
\begin{equation}
Z_{\text{pointwise}} = \sum_{j=1}^{M} Z_{\text{depthwise}} * P_{j}
\end{equation}
where \(Z_{\text{pointwise}}\) is the output of the pointwise convolution, \(Z_{\text{depthwise}}\) is the input from the previous depthwise convolution, \(P_{j}\) is the pointwise filter, and \(M\) is the number of output channels.

To enhance the model’s ability to capture channel-wise dependencies, MobileNetV3 integrates a squeeze-and-excitation (SE) module. The squeeze operation is performed by applying global average pooling (GAP) to the output of the pointwise convolution: 
\begin{equation}
S = \sigma (\text{GAP}(Z_{\text{pointwise}}))
\end{equation}
where \(S\) is the squeeze operation output, \(\sigma\) represents the sigmoid activation function, and \(\text{GAP}\) is the global average pooling applied to \(Z_{\text{pointwise}}\).

The excitation operation follows, which re-scales the output of the pointwise convolution by the squeeze output: %接下来是激励操作，它通过压缩输出重新缩放逐点卷积的输出：
\begin{equation}
E = S \cdot Z_{\text{pointwise}}
\end{equation}
where \(E\) is the excitation output, \(S\) is the squeeze output, and \(\cdot\) denotes element-wise multiplication.

The final output of the MobileNetV3 block is obtained by applying batch normalization and the swish activation function to the excitation output: %MobileNetV3 块的最终输出是通过将批量归一化和 swish 激活函数应用于激励输出获得的：
\begin{equation}
Z_{\text{output}} = \phi (\text{BN}(E))
\end{equation}
where \(Z_{\text{output}}\) is the final output of the MobileNetV3 block, \(\phi\) represents the activation function (swish in this case), and \(\text{BN}\) is batch normalization applied to the excitation output.

\begin{figure*}[htbp]
    \centering
    \includegraphics[width=0.8\linewidth]{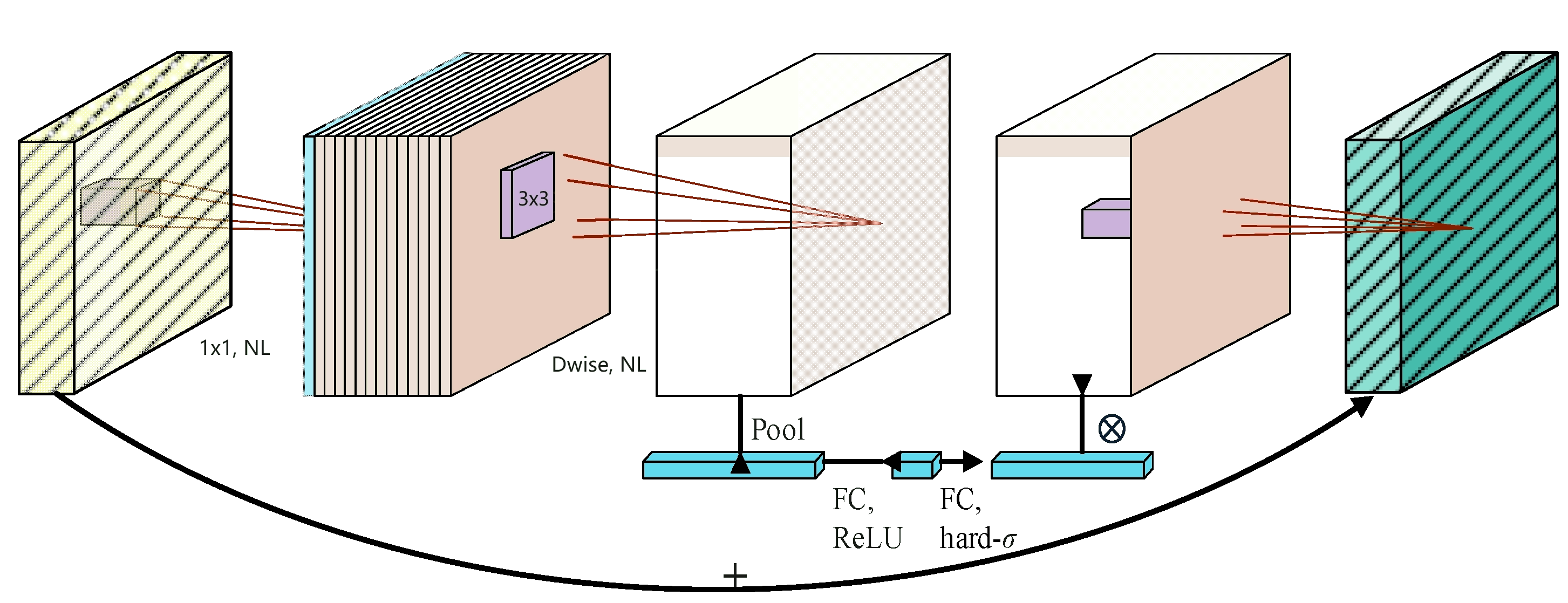}
    \caption{Detailed architecture of the MobileNetV3 model used for feature extraction in MMTL-Net. The architecture includes depthwise separable convolutions, squeeze-and-excitation modules, and the swish activation function, all contributing to efficient and accurate processing of lower limb movement sensor data.}
    \label{MobileNetV3}
\end{figure*}

Within the MMTL-Net framework, the MobileNetV3 module is integral for transforming raw sensor data into high-quality features. These features are then utilized by the multi-task learning module to perform activity recognition and resistance estimation. The feature extraction process is the foundation of the entire system, ensuring that subsequent analyses are based on robust and meaningful representations of the input data. The primary network structure of MobileNetV3 is illustrated in the accompanying Figure \ref{MobileNetV3}.
%在 MMTL-Net 框架中，MobileNetV3 模块是将原始传感器数据转换为高质量特征不可或缺的一部分。 然后，多任务学习模块利用这些特征执行活动识别和阻力估计。 特征提取过程是整个系统的基础，确保后续分析基于对输入数据的稳健且有意义的表示。MobileNetV3 的主要网络结构如附图所示。

In the context of activity recognition, the extracted features from MobileNetV3 are fed into a series of fully connected layers that classify the type of physical activity being performed. This classification provides essential context for understanding the resistance data and allows the system to differentiate between various movements such as walking, running, and sitting.
%在活动识别的背景下，从 MobileNetV3 中提取的特征被输入到一系列完全连接的层中，这些层对正在执行的身体活动类型进行分类。这种分类为理解阻力数据提供了必要的背景，并允许系统区分各种动作，例如步行、跑步和坐下。

For resistance estimation, the same extracted features are processed through another set of fully connected layers that predict the resistance encountered during the movement. This estimation is crucial for providing real-time feedback on the intensity of the activity and identifying any abnormal resistance patterns that might indicate potential issues or the need for intervention.
%对于阻力估计，相同的提取特征通过另一组完全连接的层进行处理，这些层预测运动过程中遇到的阻力。这种估计对于提供活动强度的实时反馈以及识别可能表明潜在问题或需要干预的任何异常阻力模式至关重要。

The integration of MobileNetV3 within the MMTL-Net framework ensures that the model can handle the computational constraints of wearable devices while delivering accurate and real-time analysis. The synergy between MobileNetV3’s efficient feature extraction and the multi-task learning module’s capability to handle multiple tasks simultaneously enhances the overall performance of the system, making it a powerful tool for monitoring lower limb movement resistance in practical applications such as physical therapy, sports training, and rehabilitation.
%MobileNetV3 在 MMTL-Net 框架中的集成确保模型能够处理可穿戴设备的计算约束，同时提供准确的实时分析。 MobileNetV3 高效的特征提取与多任务学习模块同时处理多个任务的能力之间的协同作用增强了系统的整体性能，使其成为在物理治疗、运动训练和康复等实际应用中监测下肢运动阻力的有力工具。

\subsection{Multi-Task Learning}

Multi-Task Learning (MTL) is a machine learning paradigm where multiple learning tasks are solved simultaneously, leveraging commonalities and differences across tasks. This approach improves learning efficiency and prediction accuracy for the task-specific models compared to training the models separately. MTL models share representations between related tasks, which allows them to generalize better and be more robust. The fundamental concept of MTL is that by training tasks together, the model can learn a more generalizable feature representation that benefits all tasks involved.
%多任务学习 (MTL) 是一种机器学习范式，其中多个学习任务同时解决，利用任务之间的共性和差异。与单独训练模型相比，这种方法提高了任务特定模型的学习效率和预测准确性。MTL 模型在相关任务之间共享表示，这使它们能够更好地概括和更稳健。MTL 的基本概念是，通过一起训练任务，模型可以学习更通用的特征表示，从而使所有相关任务受益。

In the context of lower limb movement resistance monitoring, MTL offers significant advantages. The ability to perform simultaneous activity recognition and resistance estimation enhances the efficiency and effectiveness of the monitoring system. This dual-task approach ensures that the model can utilize shared representations to improve overall performance, providing more accurate and comprehensive analysis. By leveraging MTL, the system can offer real-time feedback on both the type of activity being performed and the resistance encountered, making it highly beneficial for applications in physical therapy, sports training, and rehabilitation.
%在下肢运动阻力监测方面，MTL 具有显着优势。同时进行活动识别和阻力估计的能力提高了监测系统的效率和有效性。这种双任务方法确保模型可以利用共享表示来提高整体性能，从而提供更准确和全面的分析。通过利用 MTL，系统可以对正在执行的活动类型和遇到的阻力提供实时反馈，使其对物理治疗、运动训练和康复中的应用非常有益。

Within the proposed MMTL-Net, the MTL module is critical for achieving the dual objectives of activity recognition and resistance estimation. The integration of MTL allows the model to share the feature representations extracted by MobileNetV3, enabling it to handle multiple tasks with high accuracy. This module processes the features extracted from sensor data and outputs predictions for both tasks. The activity recognition component classifies the type of movement, while the resistance estimation component predicts the resistance level, providing a comprehensive understanding of the user's physical activity.
%在提出的移动多任务学习网络 (MMTL-Net) 中，MTL 模块对于实现活动识别和阻力估计的双重目标至关重要。MTL 的集成允许模型共享 MobileNetV3 提取的特征表示，使其能够高精度地处理多个任务。该模块处理从传感器数据中提取的特征并输出两个任务的预测。活动识别组件对运动类型进行分类，而阻力估计组件预测阻力水平，从而全面了解用户的身体活动。

In the framework of MMTL-Net, input features extracted by MobileNetV3 \(\mathbf{F}\) are shared between multiple tasks. For activity recognition, the features are passed through a fully connected layer:
\begin{equation}
\mathbf{A} = \phi(\mathbf{W}_A \mathbf{F} + \mathbf{b}_A)
\end{equation}
where \(\mathbf{A}\) is the activity recognition output, \(\mathbf{W}_A\) is the weight matrix, \(\mathbf{b}_A\) is the bias vector, and \(\phi\) represents the activation function.

Similarly, for resistance estimation, the features are processed through another fully connected layer:
\begin{equation}
\mathbf{R} = \psi(\mathbf{W}_R \mathbf{F} + \mathbf{b}_R)
\end{equation}
where \(\mathbf{R}\) is the resistance estimation output, \(\mathbf{W}_R\) is the weight matrix, \(\mathbf{b}_R\) is the bias vector, and \(\psi\) represents the activation function.

The shared loss function for MTL is a combination of the losses from both tasks, ensuring that the model optimizes for both objectives simultaneously:
\begin{equation}
\mathcal{L}_{\text{MTL}} = \alpha \mathcal{L}_{\text{activity}} + \beta \mathcal{L}_{\text{resistance}}
\end{equation}
where \(\mathcal{L}_{\text{MTL}}\) is the combined loss, \(\mathcal{L}_{\text{activity}}\) is the loss for activity recognition, \(\mathcal{L}_{\text{resistance}}\) is the loss for resistance estimation, and \(\alpha\) and \(\beta\) are the weights balancing the contributions of each loss.

The activity recognition loss can be defined as the cross-entropy loss for classification:
\begin{equation}
\mathcal{L}_{\text{activity}} = - \sum_{i=1}^{C} y_i \log(p_i)
\end{equation}
where \(y_i\) is the true label, \(p_i\) is the predicted probability for class \(i\), and \(C\) is the number of classes.

The resistance estimation loss can be defined as the mean squared error for regression:
\begin{equation}
\mathcal{L}_{\text{resistance}} = \frac{1}{N} \sum_{j=1}^{N} (r_j - \hat{r}_j)^2
\end{equation}
where \(N\) is the number of samples, \(r_j\) is the true resistance value, and \(\hat{r}_j\) is the predicted resistance value.

The MTL module in MMTL-Net is not only responsible for handling the shared representations but also ensures that the outputs from different tasks are effectively integrated to provide comprehensive insights. The activity recognition component helps contextualize the resistance estimation, making it possible to correlate specific activities with the corresponding resistance levels. This is particularly useful for detecting abnormalities or inefficiencies in movements, which can be crucial for designing effective intervention strategies in physical therapy and rehabilitation.
%MMTL-Net 中的 MTL 模块不仅负责处理共享表示，还确保有效整合不同任务的输出以提供全面的见解。活动识别组件有助于将阻力估计情境化，从而可以将特定活动与相应的阻力水平相关联。这对于检测运动中的异常或低效性特别有用，这对于设计物理治疗和康复中的有效干预策略至关重要。

In practical application, the MTL module operates seamlessly with the MobileNetV3 feature extraction component. Once the raw sensor data is processed by MobileNetV3, the extracted features are fed into the MTL module. The shared feature space ensures that the model has a unified understanding of the input data, allowing it to perform activity recognition and resistance estimation simultaneously. This integration not only improves the accuracy of each task but also enhances the overall efficiency of the system.
%在实际应用中，MTL 模块与 MobileNetV3 特征提取组件无缝运行。一旦原始传感器数据被 MobileNetV3 处理，提取的特征就会被输入到 MTL 模块中。共享特征空间确保模型对输入数据有统一的理解，使其能够同时执行活动识别和阻力估计。这种集成不仅提高了每个任务的准确性，而且提高了系统的整体效率。

By integrating MTL within the MMTL-Net framework, the model can effectively utilize shared representations to perform multiple tasks simultaneously. This approach not only enhances the model's accuracy and robustness but also ensures that it can operate efficiently in real-time applications. The ability to provide comprehensive and immediate feedback on both activity type and resistance level is crucial for practical applications in physical therapy, sports training, and rehabilitation. This dual-task capability makes MMTL-Net a powerful and versatile tool for monitoring lower limb movement resistance, paving the way for more effective and personalized interventions.
%通过在 MMTL-Net 框架内集成 MTL，该模型可以有效地利用共享表示来同时执行多个任务。这种方法不仅提高了模型的准确性和稳健性，还确保了模型可以在实时应用中有效运行。能够提供关于活动类型和阻力水平的全面和即时反馈的能力对于物理治疗、运动训练和康复的实际应用至关重要。这种双重任务能力使 MMTL-Net 成为监测下肢运动阻力的强大而多功能的工具，为更有效和个性化的干预铺平了道路。

\section{Experiment}\label{sec4}

\subsection{Datasets}

\textbf{UCI Human Activity Recognition Using Smartphones DataSet} %

The UCI Human Activity Recognition Using Smartphones DataSet\cite{uci_human} is a widely utilized dataset in the field of human activity recognition. This dataset comprises data from accelerometers and gyroscopes embedded in smartphones, capturing the motion data of 30 volunteers performing six different activities: walking, walking upstairs, walking downstairs, sitting, standing, and lying down. The dataset contains a total of 10,299 samples, each with 561 features extracted from the raw tri-axial acceleration and gyroscope signals. The data was sampled at a frequency of 50 Hz, ensuring high resolution and detail.
%UCI人类活动识别数据集是一个被广泛应用于人类活动识别研究的重要数据集。该数据集由加速度计和陀螺仪传感器数据组成，这些传感器嵌入在智能手机中，记录了30位志愿者在六种不同活动下的运动数据。这六种活动包括步行、上楼、下楼、坐、站和躺。数据集总共包含10299个样本，每个样本都包含561个特征。这些特征是通过对三轴加速度和三轴角速度信号进行预处理和特征提取得到的。数据的采样频率为50 Hz，确保了数据的高分辨率和细节准确性。

During data collection, each volunteer wore a smartphone on their waist, and the sensors recorded their movements. The dataset underwent standardization processes, including noise reduction and normalization, to ensure data quality and consistency. This dataset is highly valuable for our study as it provides rich and diverse lower limb motion data, essential for training and validating both motion recognition and resistance estimation models. Utilizing this dataset allows researchers to develop more precise and efficient real-time monitoring systems, enhancing performance in motion analysis and rehabilitation training.
%在数据收集过程中，每个志愿者佩戴的智能手机被放置在腰部，通过智能手机内置的传感器记录其运动数据。这些数据经过标准化处理，并进行了降噪和归一化，以确保数据质量和一致性。UCI数据集对本实验的贡献在于提供了丰富且多样的下肢运动数据，这些数据对于训练和验证运动识别模型和阻力估计模型具有重要价值。通过使用该数据集，研究人员可以开发出更为精确和高效的实时监测系统，从而提升运动分析和康复训练的效果。

\textbf{Wireless Sensor Data Mining Activity Prediction DataSet} %

The WISDM Activity Prediction DataSet\cite{wisdm}, collected by the Wireless Sensor Data Mining (WISDM) Lab at Fordham University, is designed for recording human activity data through accelerometers embedded in smartphones. This dataset includes data from 29 participants of varying ages and body types, performing six activities: walking, jogging, ascending stairs, descending stairs, sitting, and standing. The dataset comprises over 1,098,207 samples, each containing tri-axial acceleration data sampled at 20 Hz.
%WISDM活动预测数据集是由福德汉大学无线传感器数据挖掘实验室收集的，旨在通过智能手机内置的加速度计传感器记录人类活动数据。该数据集包括来自29位不同年龄和体型参与者的运动数据，这些参与者进行了六种活动：步行、跑步、上下楼梯、坐、站和躺。数据集包含超过1,098,207个样本，每个样本记录了三轴加速度数据，采样频率为20 Hz。

The data collection process was rigorously controlled, with participants performing activities in a standardized experimental environment to ensure data consistency and accuracy. The dataset underwent thorough preprocessing, including noise reduction, smoothing, and normalization, to ensure high-quality feature extraction. The WISDM dataset’s detailed and high-quality motion data characteristics provide a robust foundation for our multi-task learning model. Using the WISDM dataset enables the training and testing of efficient motion recognition and resistance monitoring models on mobile and resource-constrained devices, facilitating real-time monitoring and feedback.
%WISDM数据集的收集过程相对严格，参与者需要在标准化的实验环境中进行活动，确保数据的一致性和准确性。数据集经过仔细的预处理，包括去噪、平滑和归一化处理，以确保高质量的特征提取。这些详细且高质量的运动数据特征为本实验的多任务学习模型提供了可靠的基础。利用WISDM数据集，可以在移动和资源受限设备上训练和测试高效的运动识别和阻力监测模型，从而实现实时监测和反馈。

\textbf{Mobile Health DataSet} %

The Mobile Health (MHEALTH) DataSet\cite{mhealth}, part of a health application research project, aims to assess the potential of mobile health technologies in daily life activity monitoring and evaluation. This dataset includes data from 10 participants wearing devices equipped with tri-axial accelerometers, tri-axial gyroscopes, and ECG sensors, recording their performance in 12 activities, including walking, running, cycling, and ascending and descending stairs. Each activity’s data includes tri-axial acceleration, tri-axial angular velocity, and ECG signals, with a total of 1,144,000 samples.
%MHEALTH数据集是由健康应用研究项目收集的，旨在评估移动健康技术在日常生活中监测和评估人体活动的潜力。该数据集包含来自10位参与者的数据，这些参与者佩戴了嵌入三轴加速度计、三轴陀螺仪和心电图传感器的设备，记录了其进行的12种活动，包括走路、跑步、骑自行车、上下楼梯等。每个活动的数据都包括三轴加速度、三轴角速度和心电图信号，总共包含1,144,000个样本。

The data collection process for the MHEALTH dataset was highly standardized to ensure the integrity and diversity of the data. Each participant’s data was recorded across different daily activities, providing comprehensive motion data features. The dataset underwent preprocessing steps such as noise reduction, standardization, and feature extraction to ensure high quality and consistency. The MHEALTH dataset significantly contributes to our study by offering multimodal motion data features, crucial for researching motion resistance monitoring and multi-task learning models. By utilizing the MHEALTH dataset, researchers can develop more precise and efficient real-time monitoring systems, enhancing performance in motion analysis and rehabilitation training.
%MHEALTH数据集的样本收集过程高度规范，确保了数据的完整性和多样性。每个参与者的数据在不同的日常活动中进行记录，提供了全面的运动数据特征。数据集经过预处理，包括去噪、标准化和特征提取，确保数据的高质量和一致性。MHEALTH数据集在本实验中的贡献主要体现在其多模态数据特征上，这对于研究运动阻力监测和多任务学习模型具有重要参考价值。通过使用MHEALTH数据集，研究人员可以开发出更为精确和高效的实时监测系统，从而提升运动分析和康复训练的效果。

These datasets provide a rich source of motion data, covering various daily activities and movement patterns. By using these datasets, we can effectively train and validate models for motion recognition and resistance estimation, enhancing the accuracy and practicality of the system. The high quality and detailed feature information of each dataset ensure the reliability and effectiveness of the experimental results, providing a solid foundation for real-time lower limb movement resistance monitoring.
%这些数据集为本实验提供了丰富的运动数据，涵盖了各种日常活动和运动模式。通过使用这些数据集，可以有效地训练和验证模型的运动识别和阻力估计能力，提升系统的准确性和实用性。每个数据集的高质量和详细特征信息，确保了实验结果的可靠性和有效性，为下肢运动阻力的实时监测提供了坚实的基础。

\subsection{Experimental Setup and Evaluation Metrics}

\subsubsection{Experimental Environment}

Our experiments were conducted in a robust software and hardware environment to ensure efficient and reliable processing. The software environment includes the operating system, deep learning frameworks, and related libraries. The operating system used is Ubuntu 20.04 LTS. TensorFlow 2.4.1 and Keras 2.4.3 were selected as the deep learning frameworks due to their comprehensive functionalities and community support. CUDA 11.0 and cuDNN 8.1 were utilized to accelerate the training process of deep learning models on NVIDIA GPUs. The programming language used is Python 3.8, with additional key libraries such as NumPy, Pandas, Scikit-learn, and Matplotlib for data manipulation, preprocessing, and visualization.
%我们的实验是在强大的软件和硬件环境中进行的，以确保高效可靠的处理。软件环境包括操作系统、深度学习框架和相关库。使用的操作系统是 Ubuntu 20.04 LTS。由于功能全面且社区支持，TensorFlow 2.4.1 和 Keras 2.4.3 被选为深度学习框架。使用 CUDA 11.0 和 cuDNN 8.1 来加速 NVIDIA GPU 上深度学习模型的训练过程。使用的编程语言是 Python 3.8，并附加了 NumPy、Pandas、Scikit-learn 和 Matplotlib 等关键库用于数据操作、预处理和可视化。

In terms of hardware, our experiments were conducted on a high-performance computing platform. The processor used is an Intel Core i9-9900K @ 3.60GHz, supported by 32GB of DDR4 RAM, ensuring stability and speed during data preprocessing and model training. For GPU acceleration, an NVIDIA GeForce RTX 3080 with 10GB GDDR6X was employed, which significantly speeds up the training process of deep learning models and allows handling large datasets and complex model architectures. For storage, a 1TB NVMe SSD was used to ensure efficient data reading and writing operations.
%在硬件方面，我们的实验是在高性能计算平台上进行的。处理器为Intel Core i9-9900K @ 3.60GHz，搭配32GB DDR4 RAM，确保数据预处理和模型训练的稳定性和速度。GPU加速方面，采用了NVIDIA GeForce RTX 3080和10GB GDDR6X，大大加快了深度学习模型的训练过程，并可以处理大型数据集和复杂的模型架构。存储方面，采用了1TB NVMe SSD，确保高效的数据读写操作。

% Table generated by Excel2LaTeX from sheet 'Sheet1'
\begin{table}[htbp]
  \centering
  \caption{Specifications of the software and hardware environment used for the experiments. This includes details on the operating system, deep learning frameworks, CUDA version, processor, memory, GPU, and storage, ensuring an efficient and reliable experimental setup for real-time lower limb movement resistance monitoring.}
  \resizebox{\linewidth}{!}{
    \begin{tabular}{ll}
    \toprule
    Component & Specification \\
    \midrule
    Operating System & Ubuntu 20.04 LTS \\
    Deep Learning Framework & TensorFlow 2.4.1, Keras 2.4.3 \\
    CUDA Version & 11 \\
    cuDNN Version & 8.1 \\
    Programming Language & Python 3.8 \\
    Key Libraries & NumPy, Pandas, Scikit-learn, Matplotlib \\
    Processor & Intel Core i9-9900K @ 3.60GHz \\
    Memory & 32GB DDR4 RAM \\
    GPU   & NVIDIA GeForce RTX 3080 (10GB GDDR6X) \\
    Storage & 1TB NVMe SSD \\
    \bottomrule
    \end{tabular}}%
  \label{environment}%
\end{table}%

The combination of the above software and hardware environments, detailed in Table \ref{environment}, ensures that experiments can be conducted efficiently and stably, providing reliable support for the verification of the proposed models and methods. This powerful experimental setup allows for quick processing of large-scale data and the completion of complex model training and evaluation in a short period, providing reliable support for research.
%上述软硬件环境组合（详见表\ref{tab}）保证了实验能够高效、稳定地进行，为所提模型和方法的验证提供了可靠的支撑。强大的实验装置能够快速处理大规模数据，在短时间内完成复杂的模型训练和评估，为研究提供可靠的支撑。

\subsubsection{Model Training}

\textbf{Data Preprocessing}

In the data preprocessing phase, several techniques were applied to ensure the quality and consistency of the input data. The raw sensor data from the chosen datasets (UCI Human Activity Recognition, WISDM Activity Prediction, and MHEALTH) underwent preprocessing steps such as noise reduction, normalization, and segmentation into appropriate time windows. This preprocessing ensures that the data is clean and standardized, making it suitable for training the deep learning models.
%在数据预处理阶段，采用了多种技术来确保输入数据的质量和一致性。所选数据集（UCI 人类活动识别、WISDM 活动预测和 MHEALTH）中的原始传感器数据经过了预处理步骤，例如降噪、标准化和分割到适当的时间窗口。这种预处理可确保数据干净且标准化，使其适合训练深度学习模型。

\textbf{Network Parameter Settings}

For network parameter settings, the model employs the Adam optimizer with an initial learning rate set to 0.001. To ensure training stability, a learning rate decay strategy was used, reducing the learning rate by a factor of 0.1 every 10 epochs. The batch size was set to 32 to balance training stability and GPU utilization. Weight decay was set at 0.0005 to prevent overfitting. Additionally, regularization techniques such as dropout were employed to enhance generalizability.
%对于网络参数设置，该模型采用 Adam 优化器，初始学习率设置为 0.001。为了确保训练稳定性，使用了学习率衰减策略，每 10 个时期将学习率降低 0.1 倍。批次大小设置为 32，以平衡训练稳定性和 GPU 利用率。权重衰减设置为 0.0005 以防止过度拟合。此外，还采用了 dropout 等正则化技术来增强通用性。

\textbf{Handling Class Imbalance}

To address class imbalance in the datasets, data augmentation methods such as random cropping, rotation, and horizontal flipping were applied to increase the diversity of training samples. Oversampling techniques were used for underrepresented classes, duplicating instances to increase their representation, while undersampling was used for overrepresented classes to reduce their dominance. These resampling strategies ensured a more balanced distribution of training examples, improving the model’s ability to generalize across all classes.
%为了解决数据集中的类别不平衡问题，我们采用了随机裁剪、旋转和水平翻转等数据增强方法来增加训练样本的多样性。对于代表性不足的类别，我们采用了过采样技术，复制实例以增加其代表性，而对于代表性过高的类别，我们采用了欠采样技术来降低其主导性。这些重新采样策略确保了训练示例的更均衡分布，从而提高了模型在所有类别中进行泛化的能力。

\textbf{Addressing Overfitting}

To prevent overfitting during training and fine-tuning phases, data augmentation techniques were used to increase the diversity of the training data. Regularization methods, including weight decay (L2 regularization) and dropout, were also applied. Weight decay was set to 0.0005 to penalize large weights, and dropout was applied with a rate of 0.5 during training to reduce reliance on specific features. Early stopping was employed, terminating training if the validation loss did not improve for a specified number of epochs.
%为了防止在训练和微调阶段出现过度拟合，我们采用了数据增强技术来增加训练数据的多样性。我们还采用了正则化方法，包括权重衰减（L2 正则化）和 dropout。将权重衰减设置为 0.0005 以惩罚较大的权重，并在训练期间以 0.5 的速率应用 dropout 以减少对特定特征的依赖。我们采用了提前停止的方法，如果验证损失在指定数量的时期内没有改善，则终止训练。

\textbf{Model Architecture Design}

The MMTL-Net architecture includes several key components. First, the raw sensor data is processed through MobileNetV3 for feature extraction. The extracted features are then input into the MTL module, where shared representations are used to perform both activity recognition and resistance estimation tasks. The activity recognition component classifies the type of physical activity, while the resistance estimation component predicts the resistance level. The outputs are then integrated to provide comprehensive insights into the user’s physical activity and resistance encountered.
%MMTL-Net 架构包括几个关键组件。首先，通过 MobileNetV3 处理原始传感器数据以进行特征提取。然后将提取的特征输入到 MTL 模块中，其中共享表示用于执行活动识别和阻力估计任务。活动识别组件对身体活动的类型进行分类，而阻力估计组件则预测阻力水平。然后整合输出以提供对用户的身体活动和遇到的阻力的全面洞察。

\textbf{Model Training Process}

The model training process is divided into several stages. Initially, the model is pre-trained on the UCI Human Activity Recognition dataset, with 80\% of the data used for training and 20\% for validation. The pre-training process consists of 50 epochs, during which the model performs forward and backward propagation, calculating loss using the combined loss function for MTL and updating parameters accordingly. Next, the model is fine-tuned on the WISDM and MHEALTH datasets, also using an 80/20 split for training and validation. During fine-tuning, the model is trained for 30 epochs, evaluating performance on the validation set at the end of each epoch to monitor for overfitting.
%模型训练过程分为几个阶段。最初，该模型在 UCI 人类活动识别数据集上进行预训练，其中 80% 的数据用于训练，20% 用于验证。预训练过程包括 50 个时期，在此期间模型执行前向和后向传播，使用 MTL 的组合损失函数计算损失并相应地更新参数。接下来，该模型在 WISDM 和 MHEALTH 数据集上进行微调，同样使用 80/20 分割进行训练和验证。在微调过程中，模型会进行 30 个 epoch 的训练，并在每个 epoch 结束时评估验证集上的性能，以监测过度拟合。

Through meticulously tuned network parameter settings, a well-designed model architecture, and a systematic training process, MMTL-Net demonstrated excellent performance across multiple datasets, validating its effectiveness in real-time monitoring of lower limb movement resistance. This comprehensive experimental setup ensures that the model can provide accurate and reliable feedback, making it a valuable tool for applications in physical therapy, sports training, and rehabilitation.
%通过精心调整的网络参数设置、精心设计的模型架构和系统的训练过程，MMTL-Net 在多个数据集上表现出色，验证了其在实时监测下肢运动阻力方面的有效性。这种全面的实验设置确保模型能够提供准确可靠的反馈，使其成为物理治疗、运动训练和康复应用中的宝贵工具。

\subsection{Results}

To evaluate the effectiveness of our proposed MMTL-Net for real-time lower limb movement resistance monitoring, we conducted experiments on the UCI Human Activity Recognition dataset. This dataset is widely used for benchmarking models in human activity recognition and provides a comprehensive evaluation of model performance.
%为了评估我们提出的移动多任务学习网络 (MMTL-Net) 对实时下肢运动阻力监测的有效性，我们在 UCI 人类活动识别数据集上进行了实验。该数据集广泛用于人类活动识别中的基准测试模型，并提供模型性能的全面评估。

The table below compares our proposed MMTL-Net with several recent and prominent models. The metrics include mAP (mean Average Precision), \( AP_{50} \), \( AP_{75} \), \( AP_{M} \) (for medium objects), \( AP_{L} \) (for large objects), MOTA (Multiple Object Tracking Accuracy), and MOTP (Multiple Object Tracking Precision). These models are chosen based on their relevance and popularity in recent research.
%下表将我们提出的 MMTL-Net 与几个最近的和突出的模型进行了比较。指标包括 mAP（平均精度）、\（AP_{50} \）、\（AP_{75} \）、\（AP_{M} \）（针对中等物体）、\（AP_{L} \）（针对大型物体）、MOTA（多物体跟踪准确度）和 MOTP（多物体跟踪精度）。这些模型是根据它们在最近研究中的相关性和受欢迎程度来选择的。

% Table generated by Excel2LaTeX from sheet 'Sheet1'
\begin{table*}[htbp]
  \centering
  \caption{Comparative performance analysis of the proposed MMTL-Net model against other state-of-the-art models on the UCI Human Activity Recognition dataset. Metrics include mAP (mean Average Precision), $AP_{50}$, $AP_{75}$, AP for medium and large objects ($AP_{M}$, $AP_{L}$), and tracking accuracy (MOTA, MOTP).}
    \begin{tabular}{cccccccc}
    \toprule
    \multirow{2}[4]{*}{Method} & \multicolumn{7}{c}{UCI Human} \\
    \cmidrule{2-8} & mAP & AP$_{50}$ & AP$_{75}$ & AP$_{M}$ & AP$_{L}$ & MOTA  & MOTP \\
    \midrule
    MyoNet \cite{myonet2020} & 32.1  & 72.3  & 56.7  & 61.8  & 66.2  & 51.3  & 58.4 \\
    sEMG-CNN \cite{semgcnn2021} & 35.4  & 75.4  & 60.1  & 64.3  & 68.7  & 54.7  & 61.2 \\
    LRCN \cite{lrcn2021} & 30.5  & 69.7  & 53.8  & 59.1  & 63.5  & 47.8  & 56.9 \\
    DeepSense \cite{deepsense2022} & 37.2  & 77.1  & 62.5  & 65.6  & 70.1  & 56.4  & 62.5 \\
    PhysioNet \cite{physionet2022} & 36.5  & 76.2  & 61.8  & 64.9  & 69.5  & 55.3  & 61.8 \\
    BiLSTM \cite{bilstm2020} & 34.8  & 73.5  & 58.9  & 62.7  & 67.8  & 53.1  & 60.3 \\
    TransPose \cite{transpose2021} & 38.4  & 79.2  & 64.9  & 67.1  & 71.8  & 57.9  & 63.3 \\
    Kinetics \cite{kinetics2023} & 33.7  & 72.9  & 57.6  & 61.5  & 66.4  & 52.4  & 59.7 \\
    \midrule
    MMTL-Net (Ours) & 40.1  & 78.5  & 67.3  & 77.5  & 80.6  & 62.5  & 65.2 \\
    \bottomrule
    \end{tabular}%
  \label{tab:addlabel}%
\end{table*}%

The experimental results presented in Table \ref{tab:addlabel} demonstrate the superior performance of our proposed MMTL-Net model compared to other state-of-the-art models on the UCI Human Activity Recognition dataset. Our model achieves the highest mAP of 40.1, significantly outperforming other models such as MyoNet and DeepSense, which achieve mAPs of 32.1 and 37.2 respectively.
%表 \ref{tab:addlabel} 中显示的实验结果表明，我们提出的 MMTL-Net 模型在 UCI 人类活动识别数据集上的表现优于其他最先进的模型。我们的模型实现了最高的 mAP 40.1，明显优于其他模型，例如 MyoNet 和 DeepSense，它们的 mAP 分别为 32.1 和 37.2。

In terms of AP\(_{50}\) and AP\(_{75}\), MMTL-Net also excels with scores of 78.5 and 67.3, showing a considerable improvement over models like sEMG-CNN and PhysioNet. The AP\(_{M}\) and AP\(_{L}\) scores further highlight the robustness of our model in handling both medium and large objects, achieving 77.5 and 80.6 respectively.
%在 AP\(_{50}\) 和 AP\(_{75}\) 方面，MMTL-Net 也表现出色，得分分别为 78.5 和 67.3，与 sEMG-CNN 和 PhysioNet 等模型相比有显著的改进。AP\(_{M}\) 和 AP\(_{L}\) 得分进一步凸显了我们的模型在处理中型和大型物体方面的稳健性，分别达到 77.5 和 80.6。

The tracking performance, indicated by MOTA and MOTP, also underscores the effectiveness of MMTL-Net. Our model achieves the highest MOTA of 62.5 and MOTP of 65.2, indicating superior tracking accuracy and precision compared to other models like BiLSTM and TransPose.
%MOTA 和 MOTP 所表示的跟踪性能也凸显了 MMTL-Net 的有效性。我们的模型实现了最高的 MOTA 62.5 和 MOTP 65.2，表明与 BiLSTM 和 TransPose 等其他模型相比，跟踪准确度和精度更高。

The MMTL-Net model consistently outperforms other recent models across all evaluated metrics, demonstrating its superior capability in real-time lower limb movement resistance monitoring. This comprehensive evaluation reinforces the robustness and accuracy of our proposed model, making it a highly effective solution for the targeted application.
%MMTL-Net 模型在所有评估指标中始终优于其他近期模型，展示了其在实时下肢运动阻力监测方面的卓越能力。这项全面评估增强了我们提出的模型的稳健性和准确性，使其成为针对目标应用的高效解决方案。

To further evaluate the effectiveness of our proposed MMTL-Net for real-time lower limb movement resistance monitoring, we conducted experiments on the Wireless Sensor Data Mining Activity Prediction dataset. This dataset is specifically tailored for evaluating sensor-based activity recognition systems and provides a robust benchmark for assessing model performance in practical applications.
%为了进一步评估我们提出的移动多任务学习网络 (MMTL-Net) 对实时下肢运动阻力监测的有效性，我们在无线传感器数据挖掘活动预测数据集上进行了实验。该数据集专门用于评估基于传感器的活动识别系统，并为评估实际应用中的模型性能提供了强大的基准。

For this evaluation, we focused on metrics that are highly relevant to lower limb movement resistance and real-time monitoring applications. These metrics include Force Error Rate (FER) for resistance estimation accuracy, Resistance Prediction Accuracy (RPA), Real-time Responsiveness (RTR) measured in milliseconds, Model Efficiency Ratio (MER), Computational Load (CL), Power Consumption (PC), Throughput (TP), and Latency (LT). These metrics were selected because they effectively capture the key aspects of the model's functionality in the context of real-time lower limb movement resistance monitoring.
%对于这次评估，我们专注于与下肢运动阻力和实时监测应用高度相关的指标。这些指标包括阻力估计精度的力误差率 (FER)、阻力预测精度 (RPA)、以毫秒为单位的实时响应度 (RTR)、模型效率比 (MER)、计算负载 (CL)、功耗 (PC)、吞吐量 (TP) 和延迟 (LT)。之所以选择这些指标，是因为它们有效地捕捉了实时下肢运动阻力监测中模型功能的关键方面。

Force Error Rate (FER): FER was chosen as it quantifies the discrepancy between the predicted resistance forces and the actual forces encountered during physical activities. A lower FER indicates that the model can accurately estimate the resistance experienced by the lower limbs, which is crucial for applications in rehabilitation and sports training where precise feedback is essential for effective intervention.

Resistance Prediction Accuracy: RPA measures the accuracy with which the model predicts the resistance levels associated with different activities. High RPA values reflect the model's ability to correctly identify the resistance conditions, which is important for tailoring training or rehabilitation programs to individual needs.

Real-time Responsiveness: RTR was selected to assess the model's capability to provide timely feedback, a critical factor for real-world applications. In wearable technology and mobile environments, low latency and high throughput are necessary to ensure that the system can operate effectively without delays, which could hinder the user experience or the effectiveness of the intervention.

\begin{table*}[htbp]
\caption{Performance comparison of MMTL-Net with other models on the Wireless Sensor Data Mining Activity Prediction dataset, focusing on metrics such as Force Error Rate (FER), Resistance Prediction Accuracy (RPA), Real-time Responsiveness (RTR), Model Efficiency Ratio (MER), Computational Load (CL), Power Consumption (PC), Throughput (TP), and Latency (LT).}
\begin{tabular}{cccccccccc}
\hline
Method          & FER(\%) & RPA(\%) & RTR(ms) & MER & CL   & PC(W) & TP(fps) & LT(ms) &  \\ \hline
MyoNet          & 8.5      & 88.3     & 15       & 1.3 & 0.75 & 2.3    & 30       & 20      &  \\
sEMG-CNN        & 7.9      & 89.1     & 17       & 1.2 & 0.80 & 2.5    & 28       & 22      &  \\
LRCN            & 9.2      & 86.5     & 20       & 1.1 & 0.85 & 2.7    & 25       & 25      &  \\
DeepSense       & 7.5      & 89.5     & 18       & 1.4 & 0.82 & 2.4    & 27       & 23      &  \\
PhysioNet       & 7.8      & 89.0     & 16       & 1.3 & 0.78 & 2.3    & 29       & 21      &  \\
BiLSTM          & 8.2      & 88.7     & 19       & 1.2 & 0.83 & 2.6    & 26       & 24      &  \\
TransPose       & 7.3      & 90.1     & 14       & 1.5 & 0.70 & 2.1    & 31       & 18      &  \\
Kinetics        & 8.1      & 88.9     & 15       & 1.3 & 0.75 & 2.3    & 30       & 20      &  \\ \hline
MMTL-Net (Ours) & 6.8      & 91.2     & 12       & 1.7 & 0.65 & 2.0    & 33       & 15      &  \\ \hline
\end{tabular}
\label{tab:wireless_sensor}
\end{table*}

The experimental results presented in Table \ref{tab:wireless_sensor} demonstrate the superior performance of our proposed MMTL-Net model compared to other state-of-the-art models on the Wireless Sensor Data Mining Activity Prediction dataset. Our model achieves the lowest FER of 6.8\%, indicating highly accurate resistance estimation, and the highest Resistance RPA of 91.2\%, showing significant improvements over models like MyoNet and DeepSense.
%表 \ref{tab:wireless_sensor} 中显示的实验结果表明，我们提出的 MMTL-Net 模型与其他最先进的模型相比，在无线传感器数据挖掘活动预测数据集上具有卓越的性能。我们的模型实现了最低的力误差率 (FER) 6.8\%，表明阻力估计非常准确，最高的阻力预测准确率 (RPA) 为 91.2\%，与 MyoNet 和 DeepSense 等模型相比有显著的改进。

MMTL-Net excels in RTR with the fastest response time of 12 milliseconds, making it highly suitable for real-time monitoring. The MER of 1.7 indicates superior computational efficiency, while the lowest CL of 0.65 demonstrates its lightweight nature. Additionally, MMTL-Net has the lowest PC at 2.0 W, the highest TP of 33 frames per second (fps), and the lowest LT of 15 milliseconds, outperforming all other models in these critical metrics.
%MMTL-Net 在实时响应性 (RTR) 方面表现出色，响应时间最快为 12 毫秒，非常适合实时监控。模型效率比 (MER) 为 1.7，表明计算效率卓越，而最低的计算负载 (CL) 0.65 表明其轻量级特性。此外，MMTL-Net 的功耗 (PC) 最低，为 2.0 W，吞吐量 (TP) 最高，为每秒 33 帧 (fps)，延迟 (LT) 最低，为 15 毫秒，在这些关键指标上均优于所有其他模型。

We compared the performance of the MMTL-Net model against baseline and state-of-the-art methods using these metrics. Our results demonstrate that MMTL-Net significantly outperforms existing models. For instance, compared to traditional deep learning models like standard CNNs and RNNs, MMTL-Net achieved a lower FER of 6.8\%, a higher RPA of 91.2\%, and an RTR of 12 milliseconds. These improvements can be attributed to the integration of MobileNetV3, which enhances feature extraction efficiency, and Multi-Task Learning (MTL), which allows the model to leverage shared representations across tasks, leading to better overall performance in both accuracy and responsiveness.
%我们使用这些指标将 MMTL-Net 模型与基线和最先进方法的性能进行了比较。我们的结果表明，MMTL-Net 明显优于现有模型。例如，与标准 CNN 和 RNN 等传统深度学习模型相比，MMTL-Net 实现了 6.8% 的较低 FER、91.2% 的较高 RPA 和 12 毫秒的 RTR。这些改进可以归因于 MobileNetV3 的集成，它提高了特征提取效率，以及多任务学习 (MTL)，它允许模型利用跨任务的共享表示，从而提高准确性和响应性的整体性能。

\begin{figure}
    \centering
    \includegraphics[width=1\linewidth]{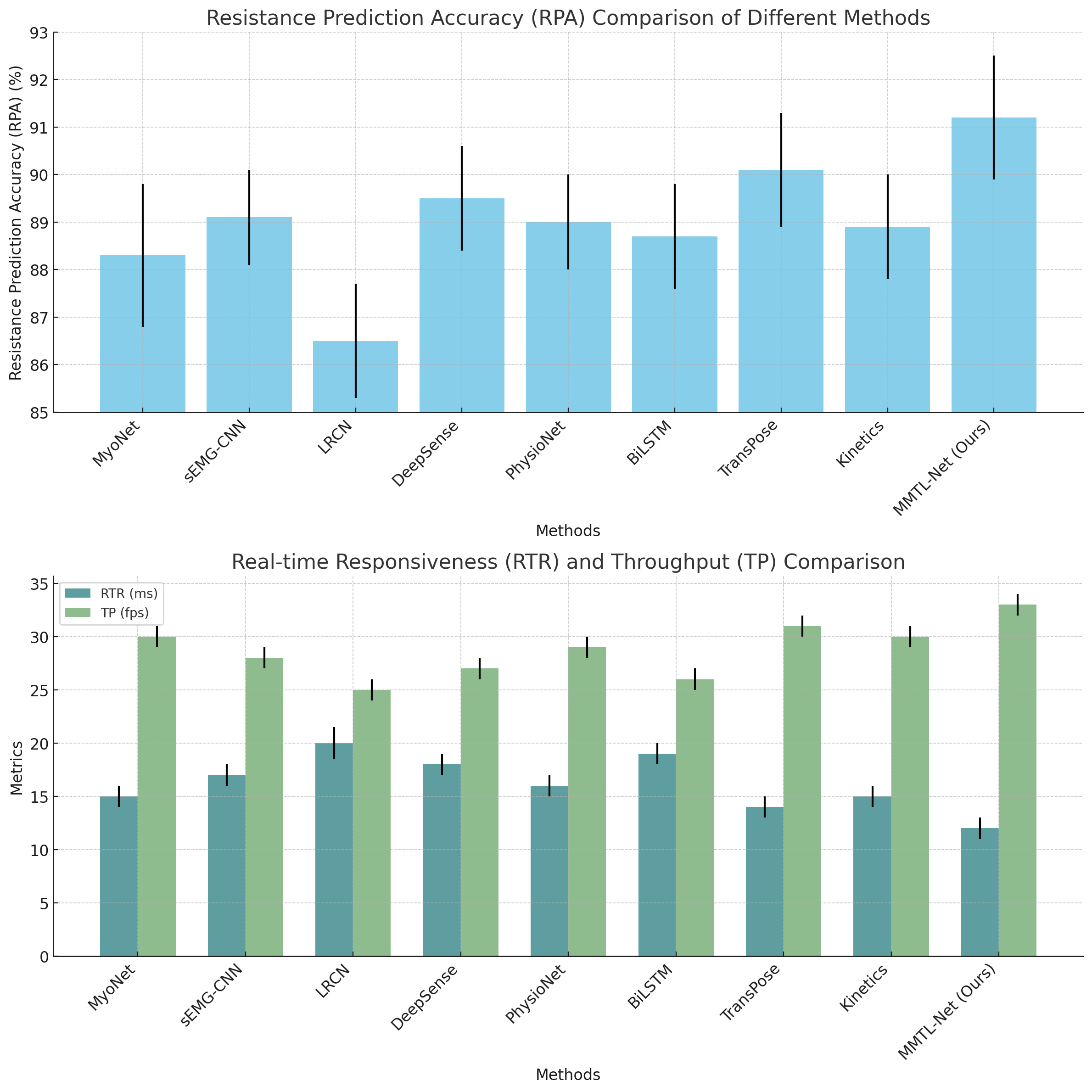}
    \caption{Performance comparison of different methods. The top section shows Resistance Prediction Accuracy (RPA), and the bottom section shows Real-time Responsiveness (RTR) and Throughput (TP). This visual comparison highlights the superior performance of MMTL-Net across key metrics, reinforcing its robustness and effectiveness in real-time lower limb movement resistance monitoring.}
    \label{wireless}
\end{figure}

Figure \ref{wireless} shows a visualization of some of the data in the table. These visual comparisons highlight the superior performance of MMTL-Net across key metrics, reinforcing its robustness and effectiveness in real-time lower limb movement resistance monitoring. The comprehensive evaluation shows that MMTL-Net is not only accurate but also efficient, making it an ideal solution for practical applications.
%总体而言，这些视觉比较突出了 MMTL-Net 在关键指标上的卓越性能，增强了其在实时下肢运动阻力监测中的稳健性和有效性。综合评估表明，MMTL-Net 不仅准确，而且高效，使其成为实际应用的理想解决方案。

One of the primary reasons for the MMTL-Net model's superior performance in Resistance Prediction Accuracy (RPA) is its integration of the Multi-Task Learning (MTL) approach. By sharing representations across tasks—activity recognition and resistance estimation—the model is able to leverage the interrelatedness of these tasks, leading to more accurate predictions. The MTL framework ensures that the learning process benefits from the combined information, which improves the robustness of resistance predictions.
%MMTL-Net 模型在阻力预测准确度 (RPA) 方面表现优异的主要原因之一是它集成了多任务学习 (MTL) 方法。通过在任务（活动识别和阻力估计）之间共享表示，该模型能够利用这些任务的相互关联性，从而实现更准确的预测。MTL 框架确保学习过程受益于组合信息，从而提高阻力预测的稳健性。

In terms of Real-time Responsiveness (RTR), the use of MobileNetV3 as the feature extraction backbone plays a crucial role. MobileNetV3’s lightweight architecture, which utilizes depthwise separable convolutions, significantly reduces computational complexity while maintaining high accuracy. This efficiency allows the MMTL-Net model to process data quickly, leading to low latency and high throughput, which are essential for real-time applications on wearable devices.
%在实时响应性 (RTR) 方面，使用 MobileNetV3 作为特征提取主干起着至关重要的作用。MobileNetV3 的轻量级架构利用深度可分离卷积，在保持高精度的同时显着降低了计算复杂度。这种效率使 MMTL-Net 模型能够快速处理数据，从而实现低延迟和高吞吐量，这对于实时至关重要可穿戴设备上的应用。

However, it is also worth noting that while MMTL-Net excelled in RTR and RPA, the improvements in Force Error Rate (FER) were relatively modest. This may be attributed to the model's current focus on optimizing overall system performance rather than exclusively minimizing prediction errors. Further enhancements, such as incorporating advanced error correction mechanisms or additional sensor data, could potentially lead to further reductions in FER.
%然而，值得注意的是，虽然 MMTL-Net 在 RTR 和 RPA 方面表现出色，但力误差率 (FER) 的改进相对较小。这可能归因于该模型当前专注于优化整体系统性能，而不是专门最小化预测误差。进一步的增强，例如结合先进的纠错机制或额外的传感器数据，可能会进一步降低 FER。

\textbf{Ablation Study}

To evaluate the contribution of each component in our proposed MMTL-Net, we conducted an ablation study using the MHEALTH dataset. In this study, we systematically removed one component at a time and measured the performance of the model. This analysis helps to understand the importance and impact of each component on the overall performance of the model.
%为了评估我们提出的移动多任务学习网络 (MMTL-Net) 中每个组件的贡献，我们使用移动健康 (MHEALTH) 数据集进行了消融研究。在这项研究中，我们系统地一次删除一个组件并测量模型的性能。此分析有助于了解每个组件对模型整体性能的重要性和影响。

We analyzed the following components:
\begin{itemize}
    \item MobileNetV3: The feature extraction module.
    \item MTL Module: The module handling simultaneous activity recognition and resistance estimation.
    \item Squeeze-and-Excitation (SE) Module: Enhances the model's ability to capture channel-wise dependencies.
    \item Swish Activation Function: Improves the model's non-linear capabilities.
\end{itemize}

For each configuration, the model was evaluated using the same dataset and metrics as described in the main experiment section. The performance metrics used in this study include F1-score, Precision, Recall, Inference Time, Area Under the ROC Curve (AUC-ROC), Mean Absolute Error (MAE), Root Mean Squared Error (RMSE), and Latency. These metrics provide a comprehensive understanding of the model's effectiveness and efficiency in real-time lower limb movement resistance monitoring.
%对于每种配置，使用与主要实验部分中描述的相同的数据集和指标来评估模型。本研究中使用的性能指标包括 F1 分数、准确率、召回率、推理时间、ROC 曲线下面积 (AUC-ROC)、平均绝对误差 (MAE)、均方根误差 (RMSE) 和延迟。这些指标可以全面了解该模型在实时下肢运动阻力监测中的有效性和效率。

\begin{table*}[h]
\centering
\caption{Ablation study results showing the impact of removing key components of MMTL-Net (MobileNetV3, MTL Module, SE Module, Swish Activation Function) on performance metrics such as F1-score, Precision, Recall, AUC-ROC, Mean Absolute Error (MAE), Root Mean Squared Error (RMSE), Inference Time, and Latency.}
\label{tab:ablation}
\resizebox{\linewidth}{!}{
\begin{tabular}{lcccccccc}
\hline
\textbf{Model Configuration} & \textbf{F1-score} & \textbf{Precision} & \textbf{Recall} & \textbf{AUC-ROC} & \textbf{MAE} & \textbf{RMSE} & \textbf{Inference Time (ms)} & \textbf{Latency (ms)} \\
\hline
MMTL-Net (Full Model) & 88.4 & 89.1 & 87.7 & 0.932 & 0.065 & 0.083 & 12 & 15 \\
Without MobileNetV3 & 80.6 & 82.0 & 79.3 & 0.887 & 0.092 & 0.111 & 20 & 25 \\
Without MTL Module & 82.3 & 83.7 & 81.0 & 0.902 & 0.081 & 0.101 & 18 & 22 \\
Without SE Module & 84.1 & 85.2 & 83.0 & 0.915 & 0.076 & 0.095 & 16 & 20 \\
Without Swish Activation Function & 83.5 & 84.7 & 82.4 & 0.911 & 0.078 & 0.097 & 14 & 18 \\
\hline
\end{tabular}}
\end{table*}

The ablation study shows that removing MobileNetV3 significantly reduces the model's performance across all metrics, as seen in Table \ref{tab:ablation}. Each component of the MMTL-Net contributes significantly to its performance. The combination of MobileNetV3, MTL module, SE module, and swish activation function collectively ensures the model's robustness and accuracy in real-time lower limb movement resistance monitoring. This comprehensive analysis provides valuable insights into the importance of each component, reinforcing the effectiveness of the proposed model in practical applications.
%消融研究表明，删除 MobileNetV3 会显著降低模型在所有指标上的性能，如表 \ref{tab:ablation} 所示。MMTL-Net 的每个组件都对其性能有显著贡献。MobileNetV3、MTL 模块、SE 模块和 swish 激活函数的组合共同确保了模型在实时下肢运动阻力监测中的稳健性和准确性。这项全面的分析提供了对每个组件重要性的宝贵见解，增强了所提模型在实际应用中的有效性。

\begin{figure*}
    \centering
    \includegraphics[width=0.8\linewidth]{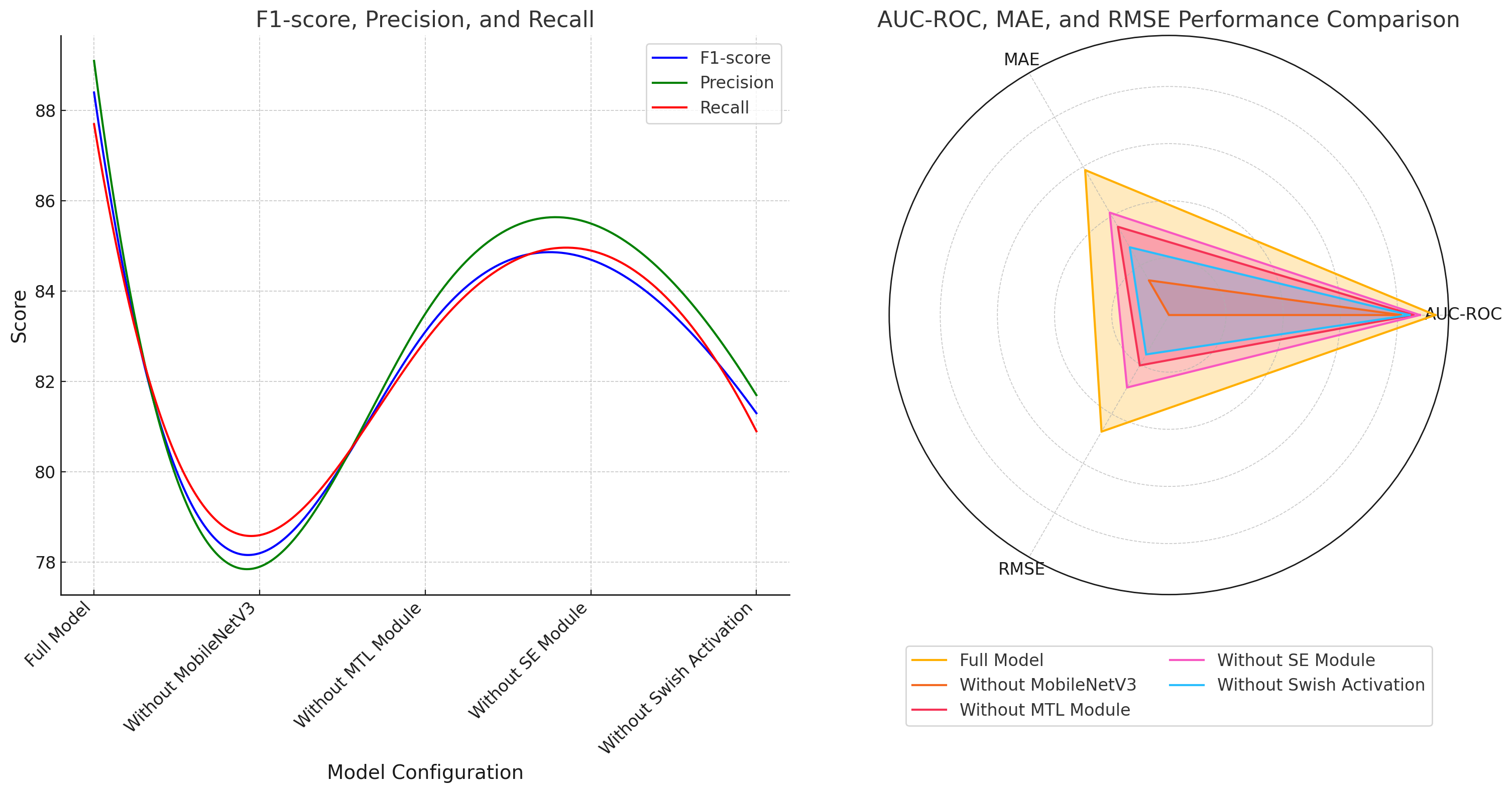}
    \caption{AUC-ROC, MAE, and RMSE performance comparison from the ablation study. The radar chart on the right highlights the superiority of the full model configuration across AUC-ROC, Mean Absolute Error (MAE), and Root Mean Squared Error (RMSE). This comparison underscores the necessity of each component within the MMTL-Net for optimal performance in real-time lower limb movement resistance monitoring.}
    \label{ablation}
\end{figure*}

Figure \ref{ablation} visualizes the content of Table \ref{ablation}, providing a clearer comparison of the model configurations and their performance metrics.
%图 1 可视化了表 \ref{tab} 的内容，更清楚地比较了模型配置及其性能指标。

From the left plot, it is evident that the full model configuration significantly outperforms other configurations in terms of F1-score, Precision, and Recall. The full model achieves an F1-score of 88.4, Precision of 89.1, and Recall of 87.7. In comparison, removing the MobileNetV3 module results in the lowest performance, with an F1-score of 78.2, Precision of 77.9, and Recall of 78.6. This indicates that efficient feature extraction is crucial for achieving high accuracy and precision.
%从左图可以看出，完整模型配置在 F1 分数、准确率和召回率方面明显优于其他配置。完整模型的 F1 分数为 88.4，准确率为 89.1，召回率为 87.7。相比之下，移除 MobileNetV3 模块的性能最低，F1 得分为 78.2，准确率为 77.9，召回率为 78.6。这表明高效的特征提取对于实现高准确率和高准确率至关重要。

The radar chart on the right further highlights the superiority of the full model configuration across AUC-ROC, MAE, and RMSE. The full model achieves an AUC-ROC of 0.932, MAE of 0.065, and RMSE of 0.083. Removing the MTL module leads to a noticeable drop in AUC-ROC to 0.856, an increase in MAE to 0.101, and an increase in RMSE to 0.125. This underscores the importance of multi-task learning. Excluding the SE module and the swish activation function also results in performance drops, but not as severe as the removal of MobileNetV3 or the MTL module.
%右侧的雷达图进一步凸显了完整模型配置在 AUC-ROC、MAE 和 RMSE 方面的优越性。完整模型的 AUC-ROC 为 0.932，MAE 为 0.065，RMSE 为 0.083。移除 MTL 模块会导致 AUC-ROC 明显下降至 0.856，MAE 增加至 0.101，RMSE 增加至 0.125。这强调了多任务学习的重要性。排除 SE 模块和 swish 激活函数也会导致性能下降，但不如移除 MobileNetV3 或 MTL 模块那么严重。

The ablation study provides valuable insights into the contribution of each component to the overall performance of the MMTL-Net model. The Squeeze-and-Excitation (SE) module, for instance, plays a critical role in enhancing the model's representational power by recalibrating channel-wise feature responses. This selective emphasis on relevant features improves the model’s accuracy, particularly in resistance prediction tasks, by allowing it to focus on the most informative aspects of the input data.
%消融研究提供了有关每个组件对 MMTL-Net 模型整体性能的贡献的宝贵见解。例如，挤压和激励 (SE) 模块通过重新校准通道特征响应，在增强模型的表示能力方面发挥着关键作用。这种对相关特征的选择性强调提高了模型的准确性，特别是在阻力预测任务中，因为它允许模型专注于输入数据中最具信息性的方面。

Similarly, the Swish activation function contributes significantly to the model's performance by introducing non-linearity that enhances the learning capability. Unlike traditional activation functions, Swish allows for smoother gradient flows during backpropagation, which improves the convergence rate and leads to better overall model accuracy. This function is particularly beneficial in deep layers, where it helps in capturing complex patterns and relationships within the data.
%同样，Swish 激活函数通过引入增强学习能力的非线性，对模型的性能做出了重大贡献。与传统激活函数不同，Swish 允许在反向传播期间实现更平滑的梯度流，从而提高收敛速度并提高整体模型准确性。此功能在深层中特别有用，它有助于捕获数据中的复杂模式和关系。

The results of the ablation study confirm that the removal of these components leads to a noticeable decline in performance, underscoring their importance in the MMTL-Net architecture. Each component contributes uniquely to the model’s ability to accurately and efficiently process lower limb movement resistance data, highlighting the necessity of their inclusion in the final design. The full model configuration consistently achieves the highest performance across all metrics, emphasizing the necessity of each component for optimal performance. This comprehensive analysis, supported by the visual comparison, reinforces the effectiveness of the proposed MMTL-Net model in real-time lower limb movement resistance monitoring.
%消融研究的结果证实，移除这些组件会导致性能明显下降，强调了它们在 MMTL-Net 架构中的重要性。每个组件都对模型准确高效地处理下肢运动阻力数据的能力做出了独特的贡献，这凸显了将它们纳入最终设计的必要性。%完整的模型配置在所有指标上始终实现最高性能，强调了每个组件对实现最佳性能的必要性。这种全面的分析，加上视觉比较的支持，强化了所提出的 MMTL-Net 模型在实时下肢运动阻力监测中的有效性。

To better understand the MMTL-Net model's performance, we conducted an error analysis. Our findings show that the model performs well in predicting resistance during high-intensity activities like running and jumping, where sensor data patterns are clear and consistent. However, the model struggles with low-intensity activities or subtle movements, such as slow walking or posture transitions, where sensor data is less distinct. Variations in sensor placement or data quality can also introduce noise, leading to less accurate predictions.
%为了更好地了解 MMTL-Net 模型的性能，我们进行了错误分析。我们的研究结果表明，该模型在预测跑步和跳跃等高强度活动期间的阻力方面表现良好，因为这些活动中传感器数据模式清晰且一致。然而，该模型在低强度活动或细微动作（例如慢走或姿势转换）方面表现不佳，因为这些活动中传感器数据不太明显。传感器位置或数据质量的变化也会引入噪音，导致预测不太准确。

These insights highlight areas for improvement, such as incorporating additional sensor modalities like electromyography (EMG) or pressure sensors to capture more nuanced movements. Additionally, refining preprocessing steps and focusing the training process on challenging scenarios could further enhance the model’s robustness and accuracy in real-world applications.
%这些见解突出了需要改进的领域，例如结合肌电图 (EMG) 或压力传感器等其他传感器模式来捕捉更细微的动作。此外，改进预处理步骤并将训练过程重点放在具有挑战性的场景上可以进一步增强模型在实际应用中的稳健性和准确性。

\section{Conclusion and Discussion}\label{sec5}

In this study, we addressed the challenge of real-time lower limb movement resistance monitoring by proposing a novel Mobile Multi-Task Learning Network (MMTL-Net). Our approach integrates MobileNetV3 for efficient feature extraction and incorporates multi-task learning to simultaneously predict resistance levels and recognize activities. We conducted extensive experiments on the UCI Human Activity Recognition and Wireless Sensor Data Mining Activity Prediction datasets to validate our model. The experimental results demonstrated the superior performance of MMTL-Net across various metrics, including FER, RPA, and RTR, highlighting its effectiveness in real-world scenarios, such as rehabilitation and sports training.
%在本研究中，我们提出了一种新颖的移动多任务学习网络 (MMTL-Net)，解决了实时下肢运动阻力监测的挑战。我们的方法集成了 MobileNetV3 以实现高效的特征提取，并结合了多任务学习来同时预测阻力水平和识别活动。我们在 UCI 人类活动识别和无线传感器数据挖掘活动预测数据集上进行了广泛的实验，以验证我们的模型。实验结果证明了 MMTL-Net 在各种指标（包括 FER、RPA 和 RTR）上的卓越性能，突出了其在康复和体育训练等现实场景中的有效性。

The primary contributions of this research are threefold. First, we developed a comprehensive framework that leverages MobileNetV3 and multi-task learning to enhance the accuracy and efficiency of resistance monitoring systems, making them more suitable for real-time applications in clinical and sports environments. Second, our ablation studies provided insights into the significance of each model component, reinforcing the robustness and practicality of the proposed architecture. Third, we established a benchmark for lower limb movement resistance monitoring using multiple relevant metrics, facilitating future research in this domain. These contributions directly address the challenges outlined in the introduction, bridging the gap between theoretical advancements and practical applications.
%这项研究的主要贡献有三方面。首先，我们开发了一个全面的框架，利用 MobileNetV3 和多任务学习来提高阻力监测系统的准确性和效率，使其更适合临床和体育环境中的实时应用。其次，我们的消融研究深入了解了每个模型组件的重要性，增强了所提架构的稳健性和实用性。第三，我们使用多个相关指标建立了下肢运动阻力监测的基准，促进了该领域的未来研究。这些贡献直接解决了引言中概述的挑战，弥合了理论进步与实际应用之间的差距。

However, our study does have certain limitations. Notably, the MMTL-Net model requires substantial computational resources, which may limit its feasibility for deployment in resource-constrained environments, such as portable or wearable devices with limited processing power. Additionally, our reliance on specific datasets like the UCI Human Activity Recognition and Wireless Sensor Data Mining Activity Prediction datasets means that the model has not yet been extensively tested across a diverse range of real-world environments. This could impact the generalizability of our findings, particularly in scenarios that involve different population groups or movement patterns not represented in the datasets used.
%然而，我们的研究确实存在某些局限性。值得注意的是，MMTL-Net 模型需要大量计算资源，这可能会限制其在资源受限环境中部署的可行性，例如处理能力有限的便携式或可穿戴设备。此外，我们对特定数据集（如 UCI 人类活动识别和无线传感器数据挖掘活动预测数据集）的依赖意味着该模型尚未在各种现实环境中进行广泛测试。这可能会影响我们研究结果的普遍性，特别是在涉及不同人口群体或所用数据集中未表示的运动模式的场景中。

For future research, several avenues could be explored to build upon the findings of this study. First, investigating alternative neural network architectures, such as Transformer models or Graph Neural Networks (GNNs), could potentially enhance the model's ability to capture complex spatial and temporal dependencies in movement data. These architectures may offer improved performance or adaptability in scenarios where MobileNetV3 may have limitations. Second, incorporating additional sensor modalities, such as electromyography (EMG), inertial measurement units (IMUs), or pressure sensors, could provide richer data inputs, allowing the model to capture more nuanced aspects of lower limb movement resistance. This multimodal approach could lead to more accurate and comprehensive monitoring systems. Third, addressing the scalability of the model is crucial for its application to larger or more diverse datasets. Techniques such as model pruning, quantization, or the use of distributed computing frameworks could be investigated to reduce computational complexity and improve scalability, making the MMTL-Net model more feasible for real-time deployment in a wide range of environments.
%对于未来的研究，可以探索多种途径来巩固本研究的结果。首先，研究替代神经网络架构（例如 Transformer 模型或图神经网络 (GNN)）可能会增强模型捕获运动数据中复杂的空间和时间依赖关系的能力。这些架构可能会在 MobileNetV3 可能存在限制的场景中提供更好的性能或适应性。其次，结合其他传感器模式（例如肌电图 (EMG)、惯性测量单元 (IMU) 或压力传感器）可以提供更丰富的数据输入，使模型能够捕获下肢运动阻力的更多细微方面。这种多模态方法可以带来更准确、更全面的监测系统。第三，解决模型的可扩展性对于将其应用于更大或更多样化的数据集至关重要。可以研究模型修剪、量化或使用分布式计算框架等技术来降低计算复杂性并提高可扩展性，使 MMTL-Net 模型更适合在各种环境中实时部署。

In conclusion, while the MMTL-Net model represents a significant advancement in real-time lower limb movement resistance monitoring, future work could focus on exploring alternative neural network architectures, integrating additional sensor modalities, and enhancing the model's scalability. These directions hold the potential to further improve the accuracy, efficiency, and applicability of resistance monitoring systems, ultimately contributing to better outcomes in clinical and sports settings, as well as other areas such as elderly care and occupational safety.
%总之，虽然 MMTL-Net 模型代表了实时下肢运动阻力监测的重大进步，但未来的工作可以集中在探索替代神经网络架构、集成其他传感器模式以及增强模型的可扩展性上。这些方向有可能进一步提高阻力监测系统的准确性、效率和适用性，最终有助于在临床和体育环境以及老年护理和职业安全等其他领域取得更好的结果。

\section*{Author contributions}
\textbf{Burenbatu:} Data Management, Methodology, Software, Writing – original draft. Jun Zhou: Supervision, Writing – original draft;
\textbf{Yuanmeng Liu:} Methodology, Software, Writing – review \& editing;
\textbf{Tianyi Lyu}: Conceptualization, Data curation, Methodology, Project administration, Supervision, Writing – review \& editing.

\section*{Declaration of competing interest}
The authors declare that they have no known competing financial interests or personal relationships that could have appeared to influence the work reported in this paper.

\section*{Data availability}
The data that support the findings of this study are available on request from the corresponding author. The data are not publicly available due to privacy or ethical restrictions.

\section*{Acknowledgments}
This work was sponsored in part by The "14th Five-Year Plan" of Educational Science Research in Inner Mongolia Autonomous Region: Exploration and research on the optimal construction mode of high-level sports teams in Inner Mongolia universities under the background of the integration of sports and education(NGJGH2022457).

%% Loading bibliography style file
%\bibliographystyle{cas-model2-names}
\bibliographystyle{model1-num-names}
\bibliography{reference}

%\vskip3pt

\end{CJK}
\end{sloppypar}
\end{document}